\documentclass[conference]{IEEEtran}
\IEEEoverridecommandlockouts 

\usepackage{times}

\usepackage{ifthen}
\newboolean{full}
\setboolean{full}{true}

\usepackage{style}

\begin{document}
\title{Coproducts of Monads on $\Set$\iffull\else\thanks{A version of this paper
    with an appendix containing detailed proofs is available at
    \texttt{www.stefan-milius.eu}}\fi}
\comment{ 
\author{\IEEEauthorblockN{\adamek\IEEEauthorrefmark{1}, 
Nathan Bowler\IEEEauthorrefmark{2},
Paul B.~Levy\IEEEauthorrefmark{3} and
Stefan Milius\IEEEauthorrefmark{1}}
\IEEEauthorblockA{\IEEEauthorrefmark{1}Institut f\"ur Theoretische
  Informatik, 
  Technische Universit\"at Braunschweig, Germany
}
\IEEEauthorblockA{\IEEEauthorrefmark{2}Fachbereich Mathematik,
  Universit\"at Hamburg, Germany}
\IEEEauthorblockA{\IEEEauthorrefmark{3}School of Computer Science,
  University of Birmingham, United Kingdom}
}
} 
\author{\IEEEauthorblockN{\adamek, Stefan Milius}
\IEEEauthorblockA{Institut f\"ur Theoretische Informatik \\ 
  Technische Universit\"at Braunschweig\\ Germany\\
\raisebox{-15pt}[0pt][0pt]{\makebox[0pt]{\hglue10.5cm\parbox[c]{\textwidth}{\begin{center}Presented at the conference  "27th Annual Symposium on Logic in Computer
Science (LICS 2012)". The current version contains proofs of some
of the results in the appendix.\end{center}}}
}}
\and
\IEEEauthorblockN{Nathan Bowler}
\IEEEauthorblockA{Fachbereich Mathematik\\
  Universit\"at Hamburg\\ Germany}
\and
\IEEEauthorblockN{Paul B.~Levy}
\IEEEauthorblockA{School of Computer Science\\
  University of Birmingham\\ United Kingdom}}

\newcommand{\theory}{\mathsf{Th}}
\newcommand{\boldS}{\mathds{S}}
\newcommand{\boldT}{\mathds{T}}
\newcommand{\boldU}{\mathds{U}}
\newcommand{\boldF}{\mathds{F}}
\newcommand{\boldM}{\mathds{M}}
\newcommand{\boldi}{\mathcal{I}}
\newcommand{\Sbar}{\bar{S}}
\newcommand{\Tbar}{\bar{T}}
\newcommand{\Ubar}{\bar{U}}
\newcommand{\injcat}{\mathsf{Inj}}
\newcommand{\cards}{\mathsf{Card}}
\newcommand{\morphism}[3]{%
   \xymatrix@1{{#2} \ar[r]^-{#1} & {#3}
   } }
\newcommand{\isomorphism}[3]{%
   \xymatrix@1{
  {#2} \ar[r]^-{#1}_{\cong} & {#3}
  } }
\newcommand{\injmorphism}[3]{%
   \xymatrix@1{
  {#2} \ar@{>->}[r]^-{#1} & {#3}
  } }
\newcommand{\longmorphism}[3]{%
   \xymatrix@1{
  {#2} \ar[rr]^-{#1} & & {#3}
  } }
\newcommand{\catc}{\mathcal{C}}
\newcommand{\eqdef}{\stackrel{\mbox{\rm {\tiny def}}}{=}}
\newcommand{\monplus}{\oplus}
\newcommand{\bigmonplus}{\bigoplus}
\newcommand{\classcard}{\mathcal{A}}
\newcommand{\oclass}{\mathcal{B}}
\newcommand{\hcone}[2]{b^{#1}_{#2}}
\newcommand{\gcone}[2]{c^{#1}_{#2}}

\maketitle

\begin{abstract}
Coproducts of monads on $\Set$ have arisen in both the study of computational effects and universal algebra.  

We describe coproducts of consistent monads on $\Set$ by an initial algebra formula, and prove also the converse: if the coproduct exists, so do the required initial algebras.  That formula was, in the case of ideal monads, also used by Ghani and Uustalu.  We deduce that coproduct embeddings of consistent monads are injective; and that a coproduct of injective monad morphisms is injective.

Two consistent monads have a coproduct iff either they have
arbitrarily large common fixpoints, or one is an exception monad,
possibly modified to preserve the empty set.  Hence a consistent monad has a coproduct with every monad iff it is an exception monad, possibly modified to preserve the empty set.  We also show other fixpoint results, including that a functor (not constant on nonempty sets) is finitary iff every sufficiently large cardinal is a fixpoint.
\end{abstract}


\section{Introduction}

The notion of \emph{monad}, in particular on the category of sets, has numerous applications.  In computer science the following two are prominent.
\begin{enumerate}
\item It is used to give semantics of \emph{computational effects}~\cite{Mo}, such as non-deterministic choice, exceptions, I/O, reading and assigning to memory cells, and control effects that capture the current continuation.
\item It provides an abstract account of the notion of ``algebraic
  theory''.  For example, a \emph{finitary} algebraic theory $\theory$ consists of a signature---a set of operations with a finite arity---and a set of equations between terms.  Then the monad $\boldT_{\theory}$ sends $X$ to the set of terms with variables drawn from $X$, modulo equivalence. 
\end{enumerate}

The \emph{coproduct} of monads $\boldS$ and $\boldT$ was studied by Kelly~\cite{Ke}, who showed that an algebra for the coproduct is a \emph{bialgebra}: a set $A$ with both an $\boldS$-algebra structure  $\sigma: SA \to A$ and a $\boldT$-algebra structure $\tau: TA \to A$.  These coproducts have arisen in both application areas:
\begin{enumerate}
\item The exception monad transformer~\cite{x}, applied to a monad $\boldT$, gives $X \mapsto T(X+E)$. This is a coproduct of $\boldT$ with the exception monad $X \mapsto X + E$.   More generally, Hyland, Plotkin and Power~\cite{HPP} gave a formula for the coproduct of a free monad $\boldF_{H}$ with a general monad $\boldT$.  This provides semantics combining I/O effects, represented by $\boldF_H$, with some other effects, represented by $\boldT$.
\item Given two theories $\theory$ and $\theory'$, we form their \emph{sum}~\cite{Pig-CM-74} by taking the disjoint union of the signatures and the union of the equation sets.  The monad $\boldT_{\theory + \theory'}$ is then a coproduct of $\boldT_{\theory}$ and $\boldT_{\theory'}$.  The sum of theories has received much attention in the field of term rewriting~\cite{BaaderTinelli}.  In particular it is shown~\cite[Prop.\ 4.14]{BaaderTinelli} that $\theory + \theory'$ is \emph{conservative} over the summands, provided each summand is consistent i.e.\ does not prove $\forall x,y.\ x=y$. This amounts to injectivity of the coproduct embeddings for the monads, and is a surprisingly nontrivial result.
\end{enumerate}
In each field some basic questions have remained.
\begin{enumerate}
\item Are there other monad transformers given by coproducts with a certain
monad $\boldT$? We give an almost negative answer: up to isomorphism, $\boldT$ must be either an exception monad or the terminal monad, possibly modified in each case to preserve the empty set.  No other monad has a coproduct with the powerset monad or with a (nontrivial) continuation monad.  This contrasts sharply with the recent result of~\cite{GS} that \emph{every} monad has a tensor with the powerset and continuation monads.
\item We can consider theories whose operations have \emph{countable} arities, or more generally arities of size $< \lambda$, for a regular cardinal $\lambda\geqslant \aleph_0$.  (Regularity ensures that, if the operations have arity $<\lambda$, then terms will too.)  These theories, and their corresponding monads, are called \emph{$\lambda$-accessible}.  Does the conservativity result hold for these?  More problematically still, there are monads, such as the powerset and continuation monads, that are not accessible (i.e.\ not $\lambda$-accessible for any $\lambda$).  We show that coproduct embeddings for consistent monads are always injective.  This subsumes the conservativity result for finitary and accessible theories.
\end{enumerate}
Kelly~\cite{Ke} showed that giving a coproduct $\boldS \monplus \boldT$ amounts to giving a free bialgebra on every set.   Three specific constructions of these coproducts appear in the literature.  Each deals in a different way with the problem of the ``shared units'': trivial terms---those that are just variables---are common to the two summands. 
\begin{figure}
  \includegraphics[scale=0.5]{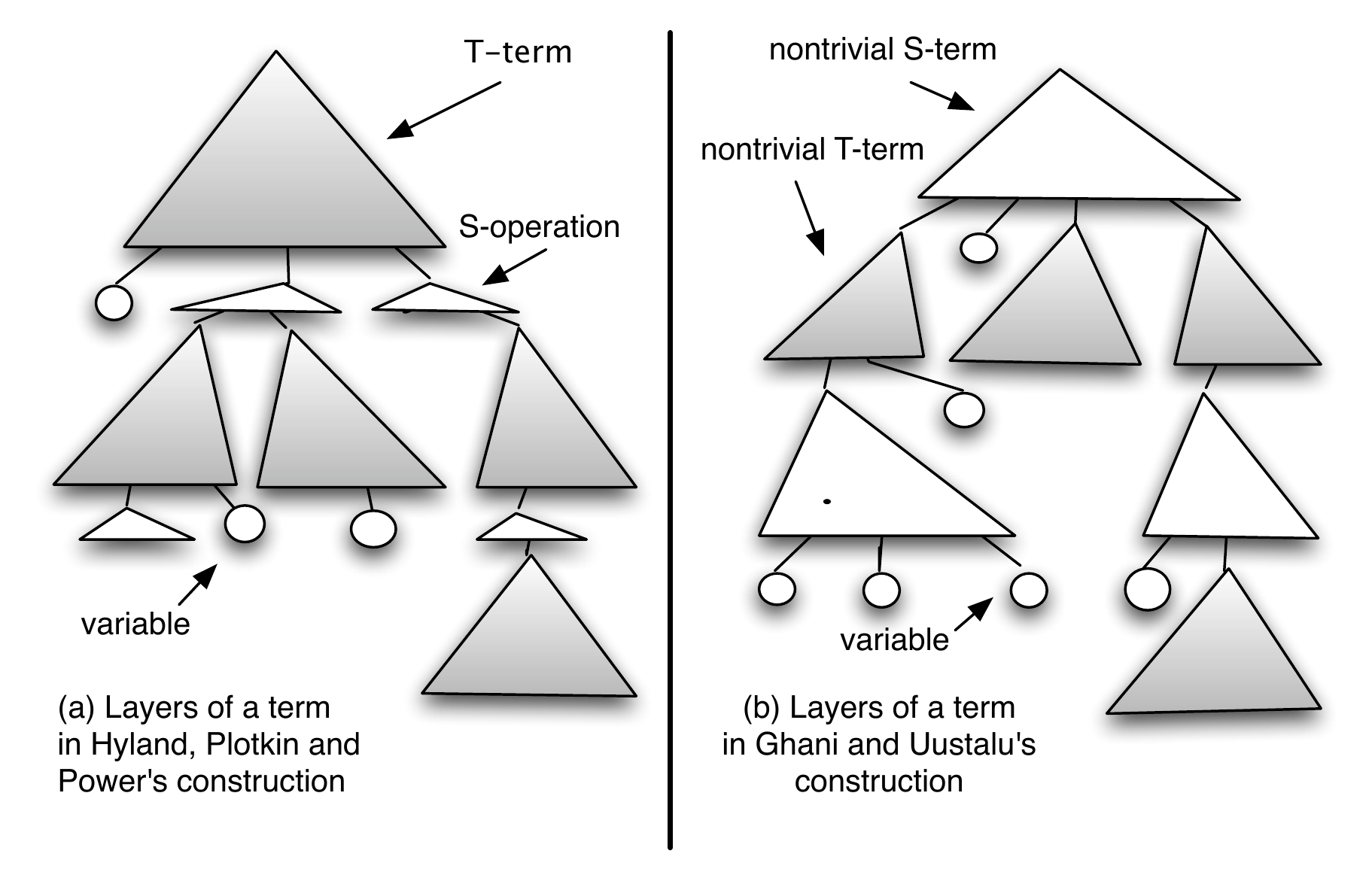}
  \caption{Layers of a term in two coproduct constructions}
  \vspace*{-10pt}
  \label{fig:layers}
\end{figure}
\begin{enumerate}[(1)]
\item Kelly~\cite{Ke} gives a multi-step construction that uses quotienting to identify the shared units.     Because it does not directly describe what gets equated, it does not enable us to prove results such as conservativity.
\item Hyland, Power and Plotkin~\cite{HPP} \label{item:hpp} treat the
  case where $\boldS$ is a \emph{free} monad, for example one arising from a theory with no equations.
Here a term in the sum consists of layers alternating between terms of $\boldT$ and operations of $\boldS$, as depicted in Fig.~\ref{fig:layers}(a), with a $\boldT$-layer uppermost. 
\item  Ghani and Uustalu~\cite{GU}\label{item:gu} treat the case where
  both $\boldS$ and $\boldT$ are \emph{ideal monads} (see Elgot~\cite{elgot}), corresponding to
  a theory whose equations are all between nontrivial terms.  A
  nontrivial term in the sum consists of layers alternating between
  nontrivial terms of $\boldS$ and those of $\boldT$, as depicted in
  Fig.~\ref{fig:layers}(b).  The uppermost layer may be of either
  kind. However, the majority of important monads, e.g.\ list, powerset, finite powerset, state and continuation monads, fail to be ideal.  
\end{enumerate}
Our first contribution is to show that Ghani and Uustalu's coproduct formula works for all consistent monads, not just ideal ones.
That seems surprising; the formula makes use of the ``ideal'', an
endofunctor on $\Set$ representing the nontrivial terms, which only an
ideal monad possesses. Our solution is to replace that ideal by the
\emph{unit complement}, an endofunctor on the category $\injcat$ of
sets and injections, possessed by every consistent monad on $\Set$, as
we shall see.

In the setting of accessible monads, the initial algebras in the
coproduct formula are guaranteed to exist, so we are done.  But in the general setting, it is only half the story: \emph{if} the initial algebras exist, we obtain a free bialgebra.  Our second contribution is to show the converse.  We therefore have a formula for a coproduct of monads whenever that coproduct exists.

This leads to our third contribution: a characterization of when a coproduct of monads exists in terms of their cardinal fixpoints: they must either have arbitrarily large common fixpoints, or else one of
them is an exception monad, possibly modified to preserve the empty set.  This has many corollaries about the existence
of coproducts between different kinds of monads.  En route we give
several new results about fixpoints, including the surprising fact
that a set functor (not constant on nonempty sets) is finitary iff every
sufficiently large cardinal is a fixpoint of it.
The last result depends on earlier work by Trnkov\'a~\cite{Tr} and
Koubek~\cite{Ko} about properties of set functors.

\noindent \textbf{Acknowledgments.}  Ohad Kammar and Gordon Plotkin
proved Lemma~\ref{lemma:injectpreserve} for finitary monads, using
Knuth-Bendix rewriting.  We thank Ohad Kammar for discussions on this
topic. The third author was supported by EPSRC Advanced Research Fellowship EP/E056091/1.

\comment{

\comment{ 
\subsection{A Problem In Universal Algebra} \label{sect:univalg}

A basic concept of algebra is a \emph{finitary theory}, i.e.\ a signature $\Sigma$ (set of operations each with a finite arity) together with a set $E$ of equations such as
\begin{equation} \label{eqn:exampleuniv}
  \forall w,x,y,z.\ f(x)*g(z) = h(z,f(z),w)
\end{equation}
Equations too have an arity, e.g.\ (\ref{eqn:exampleuniv}) has arity 4, meaning that there are 4 quantified variables.

The \emph{sum} (or disjoint union) of two theories $\theory = (\Sigma,E)$ and $\theory' = (\Sigma',E')$ is just
\begin{displaymath}
  \theory \monplus \theory' \eqdef (\Sigma + \Sigma', E \cup E')
\end{displaymath}
Is this sum always a conservative extension of $\theory$ and $\theory'$?  Using Knuth-Bendix completion, Baader and Schultz~\cite{} showed that it is, provided both $\theory$ and $\theory'$ are consistent i.e.\ do not prove $\forall x,y.\ x=y$.  Recently Kammar and Plotkin proved a more general result\footnote{Personal communication.   Their motivation was to develop modular reasoning principles for effect systems, cf.~\cite{}.   The result is more general because a theory is consistent iff its conservative restriction to the empty signature is the empty theory.}:  that for subsignatures $\Theta \subseteq \Sigma$ and $\Theta' \subseteq \Sigma'$, any equation between $\Theta+\Theta'$-terms provable in $\theory + \theory'$ is provable in $\theory\restriction_{\Theta} + \theory' \restriction_{\Theta'}$.  Here we write $\restriction$ to indicate the conservative restriction of a theory to a subsignature.

This raises several questions.  Firstly, what about theories where arities (of operations and equations) are countable rather than finite?  For example: the theory for integer state in~\cite{}.   More generally, for a regular cardinal $\lambda \geqslant \aleph_{0}$, consider \emph{$\lambda$-accessible} theories, meaning that arities are always $< \lambda$.  

Secondly, can a purely algebraic or categorical account be given?  After all, the results seem highly intuitive, and we might hope that algebraic concepts could help to bring out this intuition.  

\subsection{From Theories to Monads}

Each theory $\theory$ gives rise to a monad $\boldS_{\theory}$ on $\Set$, mapping a set to the free $\theory$-algebra on it. This erases the distinction between primitive operations and composite ones\footnote{There are several similar notions: Lawvere theory~\cite{law63}, relative monad~\cite{}, clone theory and single-object cartesian multicategory.}. As we shall explain in Sect.~\ref{sect:bk}, in the category of monads, $\boldS_{\theory + \theory'}$ is a \emph{coproduct} of $\boldS_{\theory}$ and $\boldS_{\theory'}$. 

We therefore investigate the coproduct of two monads $\boldS$ and
$\boldT$, for which 
} 

One feature of functional programs are computational effects like
non-deterministic choice, exceptions or I/O that happen on the side
while polymorphically computing something else. They are represented
by monads, see Moggi \cite{Mo}. For the semantics of a functional
programming language modularity of effects is desirable: complex
effects are obtained from simpler ones using various
constructions. Monad transformers of Cenciarelli and Moggi \cite{x}
provide such constructions; for example the exception monad
transformer $TX \to T(X+E)$. Hyland et al. \cite{HPP} noticed that this
corresponds to the formation of coproduct of the given monad
$\mathds{T}$ with the exception monad
\begin{equation*}
  \mathds{M}_E X = X+E.
\end{equation*}
Are there other monad transformers given by coproducts with a certain
monad? We give an essentially negative answer: we characterize those
monads on $\Set$ having a coproduct with
every monad. Restricted to nonempty sets such a monad is
isomorphic either to an exception monad or the trivial terminal monad.
Nevertheless, the operation
\begin{equation*}
  \mathds{S \oplus T}
\end{equation*}
of coproduct in the category of monads is interesting since it
precisely corresponds to combining the computational effects
represented by $\mathds{S}$ and $\mathds{T}$ while disregarding any
common features of these effects. Indeed, recall the observation of
Kelly \cite{Ke} that the monadic algebras for $\mathds{S \oplus T}$
are precisely the \emph{bialgebras}, that is, triples $(A, \sigma,
\tau)$ consisting of monadic algebra $\sigma: SA \to A$ for
$\mathds{S}$ and one $\tau: TA \to A$ for $\mathds{T}$. Kelly proved
that two monads have a coproduct iff every object generates a free
bialgebra. And the monad of free bialgebras is $\mathds{S \oplus T}$.
		

%
%
Three constructions of coproducts of monads appear in the literature.  Each deals in a different way with the difficulty that trivial terms---those that are just variables---are common to the two theories. 
\begin{figure}
  \includegraphics[scale=0.5]{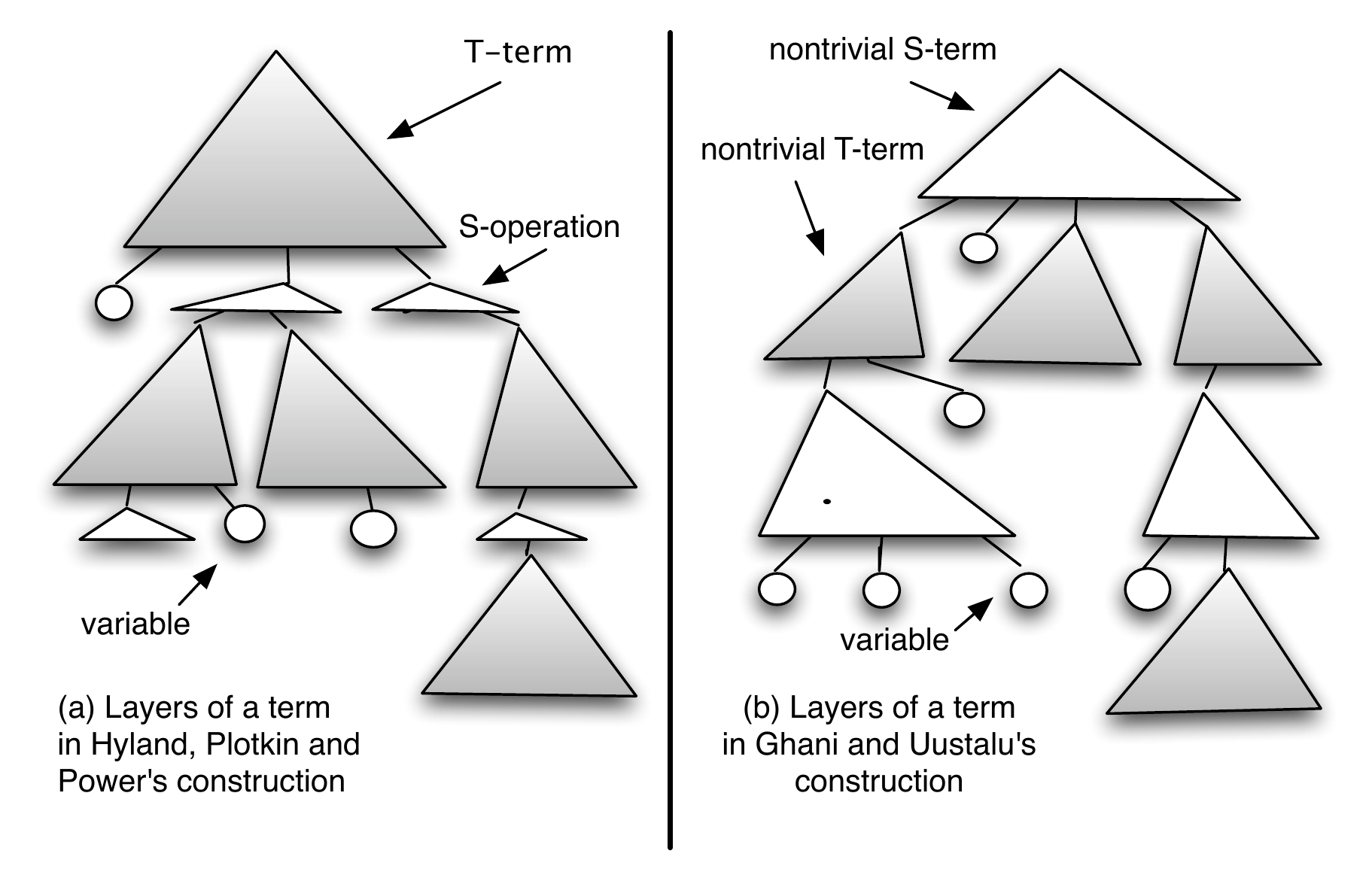}
  \caption{Layers of a term in two coproduct constructions}
  \label{fig:layers}
\end{figure}
\begin{enumerate}[(1)]
\item Hyland, Power and Plotkin~\cite{HPP} \label{item:hpp} treats the
  case where $\boldT$ is a \emph{free monad}. 
Here a term in the sum consists of layers alternating between terms of $\boldS$ and operations of $\boldT$, as depicted in Fig.~\ref{fig:layers}(a), with an $\boldS$-layer uppermost. 
\item  Ghani and Uustalu~\cite{GU}\label{item:gu} treat the case where
  both $\boldS$ and $\boldT$ are \emph{ideal monads} (see Elgot~\cite{elgot}), corresponding to
  a theory whose equations are all between nontrivial terms.  A
  nontrivial term in the sum consists of layers alternating between
  nontrivial terms of $\boldS$ and those of $\boldT$, as depicted in
  Fig.~\ref{fig:layers}(b).  The uppermost layer may be of either
  kind. This construction of $\S\oplus \T$ on objects does not involve the
  multiplication of the monads. 
\item Kelly~\cite{Ke} contains a construction of coproducts of monads which is
  quite non-trivial, involves several different steps and heavily uses the multiplication of the monads. 
\end{enumerate}
Our first contribution is to show that Ghani and Uustalu's initial
algebra formula works for all consistent monads, not just ideal ones.
This seems surprising; the formula makes use of the ``ideal'', an
endofunctor on $\Set$ representing the nontrivial terms, which only an
ideal monad possesses. Our solution is to replace that ideal by the
\emph{unit complement}, an endofunctor on the category $\injcat$ of
sets and injections, possessed by every consistent monad on $\Set$, as
we prove.  


\subsection{Accessible and Non-Accessible Monads}

A finitary theory gives a monad that preserves filtered colimits, and
conversely any such monad on $\Set$ arises from a finitary theory.  For this
reason a monad (or indeed functor) preserving filtered colimits is
called \emph{finitary}.  More generally, a functor preserving
$\lambda$-filtered colimits, for an infinite regular cardinal $\lambda \geqslant
\aleph_{0}$, is called \emph{$\lambda$-accessible}, see e.\,g.~\cite{AR}.

Accessible monads are known the have coproducts (see Kelly~\cite{Ke}).
Many important monads, such as the powerset and continuation monads,
are not $\lambda$-accessible for any $\lambda$.  Such monads do not
have coproducts in general.  Our second contribution is converse to
our first, showing that, whenever a coproduct of monads exists, it
arises from the Ghani-Uustalu formula. 

This converse leads to our third contribution: a precise
characterization of when a coproduct of monads exists: when one of
them is substantially 
an exception monad, or they have arbitrarily large common fixpoints.  

This leads in turn to an analysis of which monads have a coproduct with every monad.   Recently Goncharov and Schr\"{o}der \cite{GS} proved
that the powerset monad defines a monad transformer by tensoring:
every monad on $\Set$ has a tensor product with $\P$. 
In sharp contrast, only the substantially exceptional monads have a
coproduct with $\P$. 
		

}
\section{Universal Property of a Monad vs.\  Free Algebras} \label{sect:bk}

\begin{notation}
We write $\boldS, \boldT, \boldU$ for monads and $S, T, U$ for
endofunctors of a category $\C$, thus $\boldS = (S,\eta^{S},\mu^{S})$.  Accordingly, an ``$\boldS$-algebra'' must satisfy the Eilenberg-Moore axioms, whereas an ``$S$-algebra'' need not.
\end{notation}
\begin{remark}
  \label{R-trans}
  The \emph{transport} of an $\S$-algebra $(X, \alpha)$ along an
  isomorphism $i: X \to Y$ is the $\S$-algebra $(Y, i \o \alpha \o
  Si^{-1})$. It is easy to verify that the axioms of Eilenberg-Moore
  algebras are fulfilled. 

  In particular, given an isomorphism $i: SA \to Y$, then $Y$ is a
  free $\S$-algebra on $A$ w.r.t. the transport
  \[
  \xymatrix@1{
    SY
    \ar[r]^-{Si^{-1}} & SSA \ar[r]^-{\mu_A} & SA \ar[r]^-i & Y
    }
  \]
  and the universal arrow $i \o \eta_A: A \to Y$.
\end{remark}
In this section we review the general notions of \emph{free monad} and
\emph{coproduct of monads}. The key point is that both of these
notions have two descriptions: one using a universal property on a
monad, and one using free algebras.  Happily, on $\Set$, they turn out
to be equivalent. The proof exploits the following fact about
continuation monads $R^{(R^-)}$.
\begin{lemma}[Kelly~\cite{Ke}] \label{lemma:contbk}
  \begin{enumerate}[(1)]
  \item Let $H$ be an endofunctor on $\Set$ and $R$ a set. There is a
    bijection $\Gamma^{H}_{R}$ from $H$-algebra structures
    $HR \to R$ to natural transformations
    $H \to R^{(R^-)}$, whose inverse assigns to $\alpha\colon H
    \to R^{(R^-)}$ the algebra $\alpha_R(\id_R)\colon HR \to R$. 

  \item \label{item:contbkmonad} Let $\boldS$ be a monad on $\Set$ and
    $R$ a set.  Then $\Gamma^{S}_{R}$ gives a bijection from
    $\boldS$-algebra structures $S R \to R$ to monad
    morphisms $\boldS\to R^{(R^-)}$.
  \end{enumerate}
\end{lemma}

In the case of free monads, the two definitions are as follows.
\begin{definition}
Let $H$ be an endofunctor on a category $\catc$.
  \begin{enumerate}[(1)]
  \item A \emph{free monad} on $H$ is a monad $\boldF_{H}$ and natural transformation $\morphism{\gamma}{H}{F_{H}}$ that is initial among all such pairs $(\boldS, \morphism{\lambda}{H}{\boldS})$.
  \item Suppose every $\catc$-object $A$ generates a free
    $H$-algebra\footnote{If $\catc$ has finite coproducts, a free
      $H$-algebra on $A$ is the same thing as an initial algebra for
      $X \mapsto HX + A$.} $(F_{H}A, \rho_A)$ with unit
    $\morphism{\eta_A}{A}{F_{H}A}$.  (Equivalently: the forgetful
    functor from the $H$-algebras category to $\catc$ has a left
    adjoint.) Then the resulting monad on $\C$ is an \emph{algebraic free monad} on $H$.
  \end{enumerate}
\end{definition}

\begin{proposition}[Barr~\cite{B}] \label{prop:freemonbk}
  \begin{enumerate}[(1)]
  \item \label{item:freemonbk} Let $H$ be an endofunctor on $\catc$.
    An algebraic free monad on $H$ is a free monad with embedding
    $\gamma$ given at $A$ by
    \begin{displaymath}
      \xymatrix{
        HA \ar[r]^-{H\eta_{A}} & HF_{H}A \ar[r]^-{\rho_{A}} & F_{H}A
      }
    \end{displaymath}
  \item \label{item:freemonbkconv} Conversely, for $\catc$ with products, any free monad arises in this way.
  \end{enumerate}
\end{proposition}
\begin{corollary}
  \label{C-barr}
  For set functors $H$ the free monad $\F_H$ on $H$ fulfils $F_H A
  \cong HF_H A + A$ for every set $A$. 
\end{corollary}

In the case of coproducts, the two definitions are as follows:
\begin{definition}
  Let $\boldS$ and $\boldT$ be monads on a category $\catc$.
  \begin{enumerate}[(1)]
  \item A \emph{coproduct} of $\boldS$ and $\boldT$ is a coproduct $\boldS \monplus \boldT$ in the category of monads and monad morphisms.
  \item An \emph{$(\boldS,\boldT)$-bialgebra} $(X,\sigma,\tau)$ is an
    object $X$ with Eilenberg-Moore algebra structures
    $\morphism{\sigma}{SX}{X}$ for $\boldS$ and
    $\morphism{\tau}{TX}{X}$ for $\boldT$.
  \item Suppose that every $A \in \catc$ generates a free
    $(\boldS,\boldT)$-bialgebra $((\boldS \monplus \boldT)A,p^{S}_A,
    p^{T}_A)$ with unit $\morphism{\eta_A}{A}{(\boldS \monplus
      \boldT)A}$.  (Equivalently: the forgetful functor from the
    $(\boldS,\boldT)$-bialgebra category to $\catc$ has a left
    adjoint.)  
    Then the resulting monad is an \emph{algebraic coproduct} of $\boldS$ and $\boldT$.
  \end{enumerate}
\end{definition}

\begin{proposition}[Kelly~\cite{Ke}] \label{prop:algcoprod}
  \begin{enumerate}[(1)]
  \item \label{item:algcoprodforw} Let $\boldS$ and $\boldT$ be monads
    on $\catc$. An algebraic coproduct of $\boldS$ and $\boldT$ is $\boldS \oplus \boldT$
    with embeddings given at $A$ by
    \begin{displaymath}
\xymatrix{
      SA \ar[r]^-{S\eta_{A}} & S(\boldS \monplus \boldT)A \ar[r]^-{p^{S}_{A}} & (\boldS \monplus \boldT)A \\
            TA \ar[r]^-{T\eta_{A}} & T(\boldS \monplus \boldT)A \ar[r]^-{p^{T}_{A}} & (\boldS \monplus \boldT)A \\
}
    \end{displaymath}
  \item Conversely, for $\catc$ with products, any coproduct of monads arises in this way.
  \end{enumerate}
\end{proposition}
\comment{ 
\begin{IEEEproof}
  \begin{enumerate}[(1)]
  \item The embedding $\morphism{}{\boldS}{\boldS \monplus \boldT}$ is given at $A$ by 
    \begin{displaymath}
\xymatrix{
      SA \ar[r]^-{S \eta A} & S(\boldS \monplus \boldT) A \ar[r]^-{\sigma_{A}} & (\boldS \monplus \boldT) A
}
    \end{displaymath}
and likewise the embedding from $\boldT$, and the universal property is straightforward.
\item Similar to the proof of Proposition~\ref{prop:freemonbk}(\ref{item:freemonbkconv}), using Lemma~\ref{lemma:contbk}(\ref{item:contbkmonad}).
  \end{enumerate}
\end{IEEEproof}
} 
Thus, whilst it is  the ``algebraic coproduct'' notion that
corresponds to the joining of two theories, in $\Set$ we do not need to distinguish between the two notions.

We can easily generalize this to a coproduct of a family of monads
$(\boldS_{i})_{i \in I}$.  Here an \emph{$(\boldS_{i})$-multialgebra}
is a set $X$ with an  Eilenberg-Moore algebra structure
$\morphism{\sigma_{i}}{S_{i}X}{X}$ for $\S_i$ (where $i$ ranges
through $I$). And the monad of free $(\boldS_i)$-multialgebras is the
coproduct of the family $(\boldS_i)_{i \in I}$. 

We illustrate coproducts of monads on $\Set$ with some examples.
\begin{example}[Hyland et al.~\cite{HPP}]
  \label{E-HPP}
  We have for the exception monad $\mathds{M}_{E}: X \mapsto X +E$
  \begin{equation*}
    \boldT \oplus \mathds{M}_{E} = \boldT(- + E)
  \end{equation*}
  for all monads $\T$.
More generally, the coproduct of $\boldT$ with a family $(\boldM_{E_{p}})_{p \in P}$
of exception monads is $\boldT(- + \coprod_{p \in P} E_p)$.
\end{example}

\begin{example}
 We have, for the
  terminal monad $\mathds{1}: X \mapsto 1$
  \begin{equation*}
    \mathds{1 \oplus T = 1}
  \end{equation*}
  for all monads $\mathds{T}$. Indeed,
  $\mathds{1}$ has just one Eilenberg-Moore
  algebra (up to isomorphism), hence, there is
  only one bialgebra.  More generally, the coproduct of $\mathds{1}$
  with any family of monads is $\mathds{1}$. 
  For the submonad $\mathds{1}_0$ of the terminal monad given by $0
  \mapsto 0$ and $X \mapsto 1$ else all coproducts exist also (and are
  equal to $\mathds{1}$ or $\mathds{1}_0$). 
\end{example}

\section{Initial Algebras in $\injcat^I$}

In order to examine monads on $\Set$, we shall also have to consider categories of the form $\Set^{I}$, where $I$ is a set.  An object is an $I$-tuple of sets, often called a ``many-sorted set''.  We also need to work with $\injcat$, the category of sets and injections, and $\injcat^{I}$.   We now look at initial algebras on $\injcat^{I}$

\begin{definition}[\cite{A}]
  \label{D-chain}
Let $H$ be an endofunctor on 
a category $\C$ with colimits of chains. 
  \begin{enumerate}[(1)]\item 
  The \emph{initial
      chain} of $H$, depicted 
    \begin{displaymath}
      \xymatrix{
 0 \ar[r] & H0 \ar[r] & H^20 \ar[r] & H^30 \ar[r] & \cdots
}
    \end{displaymath}
 is a functor from $\Ord$ to $\A$ with objects $H^i0$ and connecting
    morphisms $\morphism{h_{i,j}}{H^i0}{H^j0}$ $(i \leq j$ in
    $\Ord)$. It is defined by transfinite induction on objects by
    \begin{equation*}
      H^00 = 0,\qquad H^{i+1}0 = H(H^i 0),
    \end{equation*}
    and
    \begin{equation*}
      H^i0 = \colim_{k<i} H^k0 \textrm{ for limit ordinals } i.
    \end{equation*}
    Analogously for morphisms:
    \begin{equation*}
      h_{i+1,j+1} = Hh_{i,j}
    \end{equation*}
    and for limit ordinals $i$ the cocone $(h_{k,i})_{k<i}$ is a
    colimit.  
\item The initial chain \emph{converges} at an
    ordinal $\alpha$ if the connecting map $h_{\alpha,\alpha+1}$ is invertible.
 \item For any $H$-algebra $A = (X,\theta)$, we define the \emph{canonical cocone} $(\hcone{A}{i})_{i \in \Ord}$ from the initial chain to $X$ by setting $\hcone{A}{i+1}$ to be
 \begin{displaymath}
    \xymatrix{
      H^{i+1}0 \ar[r]^-{H\hcone{A}i} & HX \ar[r]^-{\theta} & X
    }
  \end{displaymath}
  \end{enumerate}
\end{definition}

\begin{lemma} \label{lemma:initchainprop}
  Let $H$ be an endofunctor on $\injcat^{I}$. Any $H$-algebra
  homomorphism $f\colon A \to B$ 
is a morphism of canonical cocones i.e.
\[
f \o \hcone{A}{i} = \hcone{B}{i}.
\]
\end{lemma}

\begin{proposition}[Trnkov\'a et al.~\cite{TAKR}] \label{prop:initchainalg}
  Let $H$ be an endofunctor on $\injcat^{I}$.
  \begin{enumerate}[(1)]
  \item \label{item:chainconvalg} If the initial chain of $H$ converges at $i$, then the $H$-algebra  $(H^{i} 0,h_{i,i+1}^{-1})$ is initial.
  \item \label{item:algchainconv} Conversely, if there exists an $H$-algebra, then the initial chain of $H$ converges at some ordinal.
  \end{enumerate}
\end{proposition}
\begin{IEEEproof}
  \begin{enumerate}[(1)]
 \item Standard, and similar to the proof of Proposition~\ref{prop:recur} below.
\item Let $A$ be an $H$-algebra. Since $\mathsf{range}(\hcone{A}{j}) \subseteq X$ increases with $j
$, we have for some ordinal $i$ 
\begin{displaymath}
\mathsf{range}(\hcone{A}{i}) = \mathsf{range}(\hcone{A}{i+1})
\end{displaymath}
making $h_{i,i+1}$ an isomorphism.
\end{enumerate}
\end{IEEEproof}
If $H$ is finitary (i.e.\ preserves filtered colimits), then the
initial chain converges at $\omega$.  More generally, for a regular
cardinal $\lambda \geqslant \aleph_{0}$, if $H$ is
$\lambda$-accessible (i.e.\ preserves $\lambda$-filtered colimits),
then the initial chain converges at $\lambda$. 

For convenience, we shall frequently describe functors on $\injcat^{I}$, and also on $\Set^{I}$, by means of a system of equations.  For example, if $F$ and $G$ are endofunctors on $\injcat$, then an ``algebra of'' the system 
\begin{eqnarray*}
  X & = & FY \\
  Y & = & GX
\end{eqnarray*}
means an algebra for the endofunctor on $\injcat^{2}$ mapping $(X,Y)$
to $(FY,GX)$.  In this case the two components of the initial chain
take the form
\begin{equation}
  \label{eq:chain1}
  \xymatrix@1{
    0 \ar[r]
    &
    F0 \ar[r]
    &
    FG0 \ar[r]
    &
    FGF0 \ar[r]
    &
    \cdots
  }
\end{equation}
and
\begin{equation}
  \label{eq:chain2}
  \xymatrix@1{
    0 \ar[r]
    &
    G0 \ar[r]
    &
    GF0 \ar[r]
    &
    GFG0 \ar[r]
    &
    \cdots
  }
\end{equation}

We now consider the relationship between endofunctors on $\injcat^{I}$ and those on $\Set^{I}$.
\begin{definition}
  \label{D-subf}
Let $G$ be an endofunctor on $\Set^{I}$.  Suppose for each object $X$
we have a subobject $HX \subseteq GX$, in such a way that $Gm$
restricts to an injection $\injmorphism{}{HX}{HY}$ for each injection
$\injmorphism{m}{X}{Y}$.  We say that $H$ is a \emph{subfunctor on $\injcat^{I}$} of $G$.  This can be depicted as
 \begin{displaymath}
    \xymatrix{
  \injcat^{I} \ar[d]_{H} \ar@{^{(}->}[r] \ar@{}[dr]|{\subseteq} &   \Set^{I} \ar[d]^{G} \\
  \injcat^{I}  \ar@{^{(}->}[r] & \Set^{I}
}
  \end{displaymath}
\end{definition}


\begin{definition}
 Let $G$ be an endofunctor on $\Set^{I}$, with a subfunctor $H$ on $\injcat^{I}$.
  \begin{enumerate}[(1)]
  \item For an $H$-algebra $(X,\theta)$ and a $G$-algebra $(Y,\phi)$, an \emph{$H$-$G$-algebra morphism} is a function $\morphism{f}{X}{Y}$ satisfying
\begin{equation} \label{eqn:ghmorph}
  \vcenter{
\xymatrix{
 HX  \ar[r]^{\theta} \ar@{^{(}->}[d] & X \ar[dd]_{f} \\
 G X \ar[d]^{Gf} &  \\
 GY \ar[r]_{\phi} & Y
}}
\end{equation}
 \item For a $G$-algebra $A = (Y,\phi)$,  we define the
   \emph{canonical cocone} $c^A_{i}:H^i0 \to Y$ ($i \in \Ord$) from the initial chain of $H$ to $Y$ by setting $c^A_{i+1}$ to be
\begin{displaymath}
  \xymatrix{
    H^{i+1}0 \ar@{ (->}[r]^{} & GH^i0 \ar[r]^-{Gc^A_{i}} & GY \ar[r]^{\phi} & Y 
  }
\end{displaymath}
  \end{enumerate}
\end{definition}
The cocone property is established by an easy transfinite induction. 

We conclude this section by the following ``recursive function
definition'' principle.
\begin{proposition}  \label{prop:recur}
Let $G$ be an endofunctor on $\Set^{I}$, with subfunctor $H$ on $\injcat^{I}$. If $(\mu H, r)$ is an initial $H$-algebra, then for every $G$-algebra $(Y,\phi)$ there is a unique $H$-$G$-algebra morphism $\morphism{}{(\mu H,r)}{(Y,\phi)}$.
\end{proposition}
\begin{IEEEproof}
By Proposition~\ref{prop:initchainalg}(\ref{item:algchainconv}) the initial chain of $H$ converges at some ordinal $i$.  Without loss of generality we may assume $(\mu H, r) = (H^{i}0,h_{i,i+1}^{-1})$.
 For $B = (Y, \phi)$ we see that
 $\morphism{\gcone{B}{i}}{(H^{i}0,h_{i,i+1}^{-1})}{(Y,\phi)}$ is a
 homomorphism by inspecting the commutative diagram below: 
\begin{displaymath}
  \xymatrix{
 H^{i+1}0 \ar@{=}[r] \ar[d]_{h_{i,i+1}^{-1}} & H^{i+1}0 \ar@{^{(}->}[r] \ar[drr]_{\gcone{B}{i+1}} & GH^{i}0 \ar[r]^{G\gcone{B}{i}} & GY \ar[d]^{\phi} \\
 H^i0\ar[ur]_{h_{i,i+1}} \ar[rrr]_{\gcone{B}{i}} & & & Y
}
\end{displaymath}
For any $H$-$G$-algebra morphism  $f\colon A \to B$ it is easy to
prove by transfinite induction on $j \leq i$ that $f \o \gcone{A}{j} =
\gcone{B}{j}$ (cf.~Lemma~\ref{lemma:initchainprop}). For $A = (\mu H, r)$ we have $\gcone{A}{j} = h_{j,i}$,
which implies $\gcone A i = \id$. Thus, $f = \gcone{B}{i}$ is a unique
homomorphism. 
\end{IEEEproof}

\section{The Unit Complement of a Monad}

We present some basic properties of monads on $\Set$.
\begin{lemma} \label{lemma:monadinj}
Every monad $\boldS$ on $\Set$ preserves injections. 
\end{lemma}
\begin{IEEEproof}
 It suffices to show that $\morphism{\mathsf{inl}}{X}{X+Z}$ is sent to an injection.  Let $p,q\in S X$ be such that $(S\mathsf{inl})p = (S\mathsf{inl})q$.  Writing $\morphism{g}{Z}{SX}$ for the constant function to $p$,
  \begin{displaymath}
    \xymatrix{
      X \ar[r]^-{\mathsf{inl}} \ar[dr]_{\eta^{S}_{X}} & X+Z \ar[d]^{[\eta^S_X,g]} & \text{so} &  SX \ar[r]^-{S\inl} \ar[dr]^{(\eta_{X}^{S})^{*}} \ar@/_{1pc}/[dr]_{\id_{SX}} & S(X+Z) \ar[d]^{[\eta^S_X,g]^*}  \\
  & SX & & & SX
}
  \end{displaymath}
  where we write $x^*$ for $\mu_X \o Sx$. 
\end{IEEEproof}

\begin{definition}
  A monad $\boldS$ on $\Set$ is \emph{consistent} when $\eta^S_{X}$ is injective for all sets $X$. 
\end{definition}

Up to isomorphism, there are only two inconsistent monads.
\begin{lemma} \label{lemma:incon}
  If $\boldS$ is inconsistent then it is isomorphic to either $\mathds{1}$ or $\mathds{1}_{0}$.
\end{lemma}
\begin{IEEEproof}
   Suppose $\eta_X(x) = \eta_X(x')$ for some $x \not= x' \in X$. We
   show that $|SY| \leqslant 1$ for any set $Y$; hence $|SY| = 1$ if
   $Y$ is nonempty since $\morphism{\eta_X}{Y}{SY}$ cannot have empty codomain.  Given elements $p,p' \in SY$, let $\morphism{f}{X}{SY}$ be a
  function sending $x$ to $p$ and $x'$ to $p'$.  Since $f$ is
  \begin{displaymath}
    \xymatrix{
 X \ar[r]^-{\eta_{X}}  & SX \ar[r]^-{Sf} & SSY \ar[r]^-{\mu_Y} & SY
}
  \end{displaymath}
it identifies $x$ and $x'$, giving $p=p'$, so $SY = 1$.  
\end{IEEEproof}

Since we already know how to form a coproduct with $\mathds{1}$ or with $\mathds{1}_{0}$, we lose nothing by restricting attention to consistent monads.  We can then perform a fundamental construction.
\begin{definition}
  Let $\boldS$ be a consistent monad on $\Set$.  For any set $X$, we set 
  \begin{displaymath}
    \bar S X = SX \setminus \ran(\eta_X).
  \end{displaymath}
\end{definition}
In the example of a monad arising from a consistent theory, $\Sbar X$ is the set of \emph{nontrivial} equivalence classes of terms on $X$, i.e\ those classes that do not contain a variable.

\begin{proposition}
  Let $\boldS$ be a consistent monad on $\Set$. Then $\Sbar$ is a subfunctor of $\boldS$ on $\injcat$.
\end{proposition}
\begin{IEEEproof}
It suffices to show that if $p \in S X$ is sent by $S(\morphism{\mathsf{inl}}{X}{X+Y})$ into the range of $\eta^{S}_{X+Y}$ then $p \in \mathsf{range}(\eta^{S}_{X})$.  We reason as follows: either 
  \begin{eqnarray*}
    (S\mathsf{inl})p & = & (\eta^{S}_{X+Y}  \mathsf{inl}) x \\
 & = & (S\mathsf{inl}  \eta^{S}_{X})x
  \end{eqnarray*}
giving $p = \eta^{S}_{X}x$ by Lemma~\ref{lemma:monadinj}, or
\begin{equation}
(S\mathsf{inl})p = \eta^{S}_{X+Y} \mathsf{inr}\ y.
\label{eqn:unitcomplinl}
\end{equation}
In the latter case, we apply $S (\xymatrix{
 X+Y \ar[r]^-{[\mathsf{in}_{0}, \mathsf{in}_{1}]} & X+Y+Y
})$ to (\ref{eqn:unitcomplinl}) giving
\begin{displaymath}
  (S\mathsf{in}_{0}) p = \eta^{S}_{X+Y+Y} \mathsf{in}_{1} y.
\end{displaymath}
We also apply $S (\xymatrix{
 X+Y \ar[r]^-{[\mathsf{in}_{0}, \mathsf{in}_{2}]} & X+Y+Y
})$ to (\ref{eqn:unitcomplinl}) giving
\begin{displaymath}
  (S\mathsf{in}_{0}) p = \eta^{S}_{X+Y+Y} \mathsf{in}_{2} y.
\end{displaymath}
Injectivity of $\eta^{S}_{X+Y+Y}$ gives $\mathsf{in}_1 y = \mathsf{in}_2 y$, which is impossible.
\end{IEEEproof}
We call $\Sbar$ the \emph{unit complement} of $\boldS$.  
By contrast with the ``ideal monad'' framework of~\cite{GU}, $\Sbar$ might not extend to an endofunctor on $\Set$:
\begin{examples} \label{E-consi}
  \begin{enumerate}[(1)]
  \item If $\boldS$ is the finite powerset monad, then $\Sbar X$ is the set of all non-singleton finite subsets of $X$.  For the (non-injective) function $\morphism{g}{2}{1}$, we cannot define $\morphism{\Sbar g}{\Sbar 2}{\Sbar 1}$ consistently with $Sg$.
  \item If $\boldS$ is the finite list monad $X \mapsto X^\ast$,  then
    $\bar{S}X$ is the set of all words of length $\not=1$.  In this
    case $\Sbar$ \emph{does} extend to an endofunctor on $\Set$.
    Nevertheless $\boldS$ is not an ideal monad---$\mu^{S}$ does not
    map $\Sbar S$ to $\Sbar$.
  \end{enumerate}
\end{examples}

\begin{lemma} \label{lemma:barprops}
  Let $\boldS$ be a consistent monad on $\Set$. For any regular cardinal $\lambda \geqslant \aleph_{0}$, if $S$ is $\lambda$-accessible, so is $\Sbar$. 
\end{lemma}

\section{Initial Bialgebras and Multialgebras}

We saw in Sect.~\ref{sect:bk} that, to find the coproduct of two
monads $\boldS$ and $\boldT$, we need a free bialgebra on each set
$A$.  In this section, we study the simpler problem of finding an
initial bialgebra (i.e.\ $A = \emptyset$).  We shall see in
Sect.\ref{sect:gu} that this enables us to solve the general
problem. When writing $+$ we always mean coproduct in $\Set$.



To find an initial bialgebra for $\boldS$ and $\boldT$, we seek an
initial algebra in $\Inj$ for the system 
\begin{equation} \label{eqn:initbitwo}
\begin{array}{ccc}
  X & = & \Sbar Y \\
  Y & = & \Tbar X
\end{array}
\end{equation}
If it exists, we call it $(S^*,T^*)$. The algebra structure is called 
$r^S\colon \bar S T^* \stackrel{\cong}{\to} S^*$ and $r^T\colon \bar T
S^* \stackrel{\cong}{\to} T^*$. By Proposition~\ref{prop:initchainalg} this exists
whenever~\refeq{eqn:initbitwo} has a solution. This is in particular
the case if $\S$ and $\T$ are $\lambda$-accessible.

\begin{theorem} \label{thm:initbialg}
  Let $\boldS$ and $\boldT$ be consistent monads on $\Set$.
  \begin{enumerate}[(1)]
  \item \label{item:initbialgforw} If $(S^{*},T^*{})$ exists, then
    \begin{equation} \label{eqn:initbialg}
      (S^{*} + T^*, p^{S}, p^{T})
    \end{equation}
    is an initial $(\boldS,\boldT)$-bialgebra, where $p^{S}\colon S(S^* + T^*) \to
    S^* + T^*$ is the free $\boldS$-algebra on $T^*$ transported (see
    Remark~\ref{R-trans}) along the isomorphism 
    \begin{displaymath}
      ST^* \cong \longmorphism{r^{S} + T^*}{\Sbar T^* + T^*}{S^* + T^*}
    \end{displaymath}
    and $p^{T}$ is defined similarly.
  \item \label{item:initbialgconv} Conversely, any initial $(\boldS,\boldT)$-bialgebra arises in this way.
  \end{enumerate}
\end{theorem}
Explicitly, the unique bialgebra morphism from (\ref{eqn:initbialg})
to an $(\boldS,\boldT)$-bialgebra $(B,\sigma,\tau)$ is constructed as
follows.  The functor given by (\ref{eqn:initbitwo}) is a subfunctor on $\injcat^{2}$ of the functor
\begin{equation} \label{eqn:initbitwobig}
\begin{array}{ccc}
  X & = & S Y \\
  Y & = & T X
\end{array}
\end{equation}
on $\Set^2$ in the sense of Definition~\ref{D-subf}. 
Now $(B,\sigma,\tau)$ is an algebra of (\ref{eqn:initbitwobig}), so by Proposition~\ref{prop:recur}, we obtain unique $\morphism{f^S}{S^*}{B}$ and $\morphism{f^T}{T^*}{B}$ such that the squares \begin{equation} \label{EQ 3.2}
				\vcenter{
				\xymatrix{
					\bar{S}T^\ast \ar[r]^{r^S} \ar@{^{(}->}[d]
					&
					S^\ast \ar[dd]^{f^S}
					&
					\bar{T}S^\ast \ar[r]^{r^T} \ar@{^{(}->}[d]
					&
					T^\ast \ar[dd]^{f^T}
					\\
					ST^\ast \ar[d]_{Sf^T}
					&&
					TS^\ast \ar[d]_{Tf^S}
					\\
					SB \ar[r]_\sigma
					&
					B
					&
					TB \ar[r]_\tau
					&
					B
				}}
			\end{equation}
			commute.  Then the bialgebra morphism is given by
                        \begin{displaymath}
                          \longmorphism{[f^S,f^T]}{S^*+T^*}{B}.
                        \end{displaymath}
\begin{IEEEproof}[Sketch of proof]
  For~(1) we prove by diagram chasing that $[f^S, f^T]$ is a
  homomorphism for both monads $S$ and $T$. 

  For~(2), assuming that an initial bialgebra on a set $A$ is given,
  we prove that the initial chain ($S^*_i,T^*_i)$ converges by
  verifying that the canonical cocone (Definition~\ref{D-chain}) has
  all components injective from which the statement easily
  follows. The main technical trick of the proof is that for every
  sufficiently large ordinal $i$ we construct a bialgebra such that the
  canonical cocones have their components at $i$ injective.
\end{IEEEproof}

\begin{remark} 
  \label{rem:muplusmu}
  The carrier $S^* + T^*$ of the initial $(\boldS,\boldT)$-bialgebra can be written as
  \begin{displaymath}
    \mu \Sbar \Tbar + \mu \Tbar \Sbar
  \end{displaymath}
  Indeed, in the chain~\refeq{eq:chain1} all even members form the
  initial chain of $FG$, analogously with~\refeq{eq:chain2}.
\end{remark}

\begin{lemma} \label{L-initinj} Let
  $\morphism{\alpha}{\boldS}{\boldS'}$ and
  $\morphism{\beta}{\boldT}{\boldT'}$ be injective monad morphisms.
  If there is an initial $(\boldS',\boldT')$-bialgebra $(I',m',n')$,
  then there is an initial $(\boldS,\boldT)$-bialgebra $(I,m,n)$, and
  the unique $(\boldS,\boldT)$-bialgebra homomorphism
  \begin{displaymath}
    \morphism{f}{(I,m,n)}{(I',m'\o\alpha_{I'},n'\o\beta_{I'})}
  \end{displaymath}
  is injective.
\end{lemma}
\begin{remark}
  \label{rem:V.6}
  To find an initial \emph{multi}algebra for more than two monads, we have to adapt (\ref{eqn:initbitwo}).  
  \begin{itemize}\item 
    In the case of three consistent monads $\boldS,\boldT,\boldU$ we
    take in $\Inj$ the
    initial algebra $(S^*,T^*,U^*)$ of the equations
    \begin{equation}
      \begin{array}{ccc}
        X &= & \bar{S}(Y+Z)\\
        Y &= & \bar{T}(X+Z)\\
        Z &= &\bar{U}(X+Y)
      \end{array}\label{eqn:initbithree}
    \end{equation}
    and then the initial trialgebra is carried by $S^* + T^* + U^*$.
  \item In the case of a family $(\mathds{S}_p)_{p \in P}$ of
    consistent monads, we take in $\Inj$ the initial algebra $(S^*_{p})_{p \in P}$ of the equations
    \begin{equation*}
      X_p = \bar{S}_p (\sum_{q \in P \setminus \{p\}} X_q) \qquad (p \in P)
    \end{equation*}
    with structure $(r_{p})_{p \in P}$ and the initial multialgebra is carried by $\sum_{p \in P} S^*_{p}$.  The $\boldS_{p}$-structure is given by the free $\S_p$-algebra structure on
    \begin{displaymath}
      S^+_{p} \eqdef \sum_{q \in P\setminus\{p\}} S^{*}_q
    \end{displaymath}
    transported along the isomorphism
\begin{equation*}
  \sum_{p \in P} S^\ast_p \cong \xymatrix@1@C+1.5pc{S^\ast_p + S^+_p
    \ar[r]^-{r_{p}^{-1} + S^+_p} & \bar{S}_p S^+_p + S^+_p} \cong S_p(S^+_p)
\end{equation*}
\end{itemize}
\end{remark}
All the results of this section (except Remark~\ref{rem:muplusmu}) go through in this more general setting.
              
\section{Coproducts of Monads} \label{sect:gu}

In this section a formula for coproducts of monads on $\Set$ is
presented. We denote by $+$ coproducts in $\Set$ and by $\oplus$
coproducts of monads. 

\begin{remark}
  \label{R-tri}
  Suppose we have consistent monads $\boldS$ and $\boldT$, and we want
  a free $(\boldS,\boldT)$-bialgebra on a set $A$.  This is the same
  thing as an initial $(\boldS,\boldT,\boldM_{A})$-trialgebra, 
  where $M_AX = X + A$ is the exception monad,
  since an $\M_A$-algebra on $X$ corresponds to a morphism
  $A \to X$.  We know that this initial trialgebra is given
  by an initial algebra of (\ref{eqn:initbithree}), which in this case
  takes the form
\begin{eqnarray*}
  X & = & \Sbar(Y + Z) \\
  Y & = & \Tbar(X + Z) \\
  Z & = & A
\end{eqnarray*}
By an elementary argument this corresponds to an initial algebra of
\begin{equation}
\begin{array}{ccc}
  X & = & \Sbar(Y+A) \\
  Y & = & \Tbar(X +A)
\end{array}\label{eqn:freebitwo}
\end{equation}
Recall that these initial algebras are taken in $\Inj$.
\end{remark}

\begin{definition}
  Let $\boldS$ and $\boldT$ be consistent monads on $\Set$.
  \begin{enumerate}[(1)]
  \item For any set $A$, we define $(S^*A, T^*A)$ to be an initial algebra of (\ref{eqn:freebitwo}) if it
    exists.  The algebra structure is called 
    \[
    s_A^*\colon \Sbar (T^* A + A) \stackrel{\cong}{\to} S^* A
    \ \text{and}\ 
    t_A^*\colon \Tbar (S^* A + A) \stackrel{\cong}{\to} T^* A.
    \]
    
    \comment{ 
  \item If this exists for every set $A$, then we extend $S^*$ and $T^*$ to endofunctors on $\injcat$ in the unique way that makes 
    \begin{displaymath}
      \xymatrix{
 \Sbar(T^*A + A) \ar[r]^-{s^*_A} & S^*A & 
 \Tbar(S^*A  + A) \ar[r]^-{t^*_A} & T^*A \\
}
    \end{displaymath}
natural in $A$.
} 
\item Consider $S^*A + T^* A + A$ to be a bialgebra as follows. 
  Denote by $p^S_A\colon S(S^* A + T^* A + A) \to S^* A+ T^* A +
  A$ the free $\S$-algebra on $T^*A + A$ transported (see Remark~\ref{R-trans}) along the isomorphism 
  \begin{displaymath}
    \xymatrix@C-2pc{
      S(T^*A + A) & \cong  &  \Sbar(T^*A + A) + T^*A + A
      \ar[d]_{s^*_A + T^*A + A} \\
      S^*A + T^*A + A & \cong &  S^*A + (T^*A + A)
    }
    \end{displaymath}
    and $p^{T}_A$ 
    is the free $\boldT$-algebra on $S^*A + A$ transported along the
    analogous isomorphism. 
  \end{enumerate}
\end{definition}

\begin{proposition} \label{P-freebialg}
  Let $\boldS$ and $\boldT$ be consistent monads on $\Set$.  Let $A$ be a set.
  \begin{enumerate}[(1)]
  \item If $(S^{*}A,T^* A)$ exists, then
    \begin{equation} \label{eqn:freebialg}
      (S^{*}A + T^*A + A, p^{S}_A, p^{T}_A) 
    \end{equation}
    with unit $\inr\colon A \to S^* A + T^* A + A$ is a free $(\boldS,\boldT)$-bialgebra on $A$.
\item Conversely, any free $(\boldS,\boldT)$-algebra on $A$ arises in this way.
  \end{enumerate}
\end{proposition}
Explicitly, the unique bialgebra morphism from the above algebra to an $(\boldS,\boldT)$-bialgebra $(B,\sigma,\tau)$ extending $\morphism{h}{A}{B}$ is constructed as follows.  By Proposition~\ref{prop:recur}, we obtain unique $\morphism{f^S}{S^* A}{B}$ and $\morphism{f^T}{T^* A}{B}$ such that
\[
\xymatrix{
  \bar{S}(T^\ast A + A) \ar[rr]^{s^*_A}
  \ar@{^{(}->}[d] & & S^\ast A \ar[dd]^{f^S} \\
  S(T^\ast A + A) \ar[d]_{Sf^T} & & \\
  S(B + A) \ar[r]_-{S[\id,h]} & SB \ar[r]_{\sigma} & B 
}
\] 
and
\[
\xymatrix{
  \bar{T}(S^\ast A +A) \ar[rr]^{t^*_A} \ar@{^{(}->}[d] & & T^\ast A \ar[dd]^{f^T} \\
    T(S^\ast A +A)
    \ar[d]_{Tf^S} & & \\
    T(B + S)
    \ar[r]_-{T[\id,h]} & TB \ar[r]_\tau & B
  }
\]
commute.  Then the bialgebra morphism is given by
\begin{displaymath}
  \longmorphism{[f^S,f^T,h]}{S^*+T^* + A}{B}
\end{displaymath}
It is easily checked that this is the construction derived from that
in Theorem~\ref{thm:initbialg} and Remark~\ref{rem:V.6}.

\begin{theorem} \label{thm:gu}
  A coproduct of monads $\boldS$ and $\boldT$ on $\Set$ exists iff one
  of the monads is inconsistent or an initial algebra $(S^* A, T^* A)$
  for~\refeq{eqn:freebitwo}  exist in $\Inj$ for all $A$.  Under these circumstances:
  \begin{enumerate}[(1)]
  \item \label{item:guset} $(\boldS \monplus \boldT)A$ is given by $(S^*A + T^*A) + A$ for every set $A$
  \item \label{item:guunit} the unit of $\boldS \monplus \boldT$ is given at $A$ by 
    \begin{displaymath}
\xymatrix{
      A \ar[r]^-{\mathsf{inr}} & S^*A + T^*A + A
}
    \end{displaymath}
  \end{enumerate}
\end{theorem}
\begin{remark}
  The coproduct embedding $\morphism{}{\boldS}{\boldS \monplus \boldT}$ is given at $A$ by
   \begin{displaymath}
     \xymatrix{
       SA \cong 
       \Sbar A + A \ar[r]^-{\Sbar \mathsf{inr} + A} 
       & 
       \Sbar(T^*A + A) + A  \ar[d]^{s^*_A + A} \\
       S^*A + T^*A + A  & S^{*}A + A \ar[l]^-{\mathsf{inl} + A}
     }
   \end{displaymath}
   and likewise for the embedding $\morphism{}{\boldT}{\boldS \monplus \boldT}$. 
\end{remark}

\begin{corollary}\label{cor:embedinj}
 If $\boldS$ and $\boldT$ are consistent monads and $\boldS \monplus \boldT$ exists, then $\boldS \monplus \boldT$ is consistent and the coproduct embeddings
  \begin{displaymath}
    \xymatrix{
      \boldS \ar[r]^-{\mathsf{inl}} & \boldS \monplus \boldT & \boldT \ar[l]_-{\mathsf{inr}}
    } 
  \end{displaymath}
  are injective.
\end{corollary}

\begin{lemma} \label{lemma:injectpreserve}
  Let $\boldS',\boldT'$ be consistent monads such that $\boldS'
  \monplus \boldT'$ exists.  For any injective monad morphisms
  $\morphism{i}{\boldS}{\boldS'}$ and $\morphism{j}{\boldT}{\boldT'}$
  \begin{itemize}
  \item $\boldS \monplus \boldT$ exists
  \item the monad morphism $\morphism{i \oplus j}{\boldS \monplus \boldT}{ \boldS' \monplus \boldT'}$ is injective.
  \end{itemize}
\end{lemma}
\begin{IEEEproof}
  Analogous to Remark~\ref{R-tri}, for each set $A$, the initial
  $(\S', \T', \M_A)$-trialgebra $(I',m',n',a')$ exists.  Therefore by
  Lemma~\ref{L-initinj} the initial trialgebra $(I,m,n,a)$ of
  $\boldS$, $\boldT$ and $\boldM_{A}$, exists, i.e.\ the free
  $(\boldS,\boldT)$-bialgebra on $A$, giving $(\boldS \monplus
  \boldT)A$.  Moreover, Lemma~\ref{L-initinj} gives the injectivity of
  the unique trialgebra morphism from $(I,m,n,a)$ to
  $(I',m'\o\alpha_{I'},n'\o\beta_{I'}, a')$, i.e.\ the unique bialgebra
  morphism commuting with the units, which is precisely $(i \monplus j)_A$.
\end{IEEEproof}

To form the coproduct of a family $(\mathds{S}_p)_{p \in P}$ of
consistent monads, we take for each set $A$ the initial algebra
$(S^*_{p}A)_{p \in P}$ of the equations
\begin{equation*}
  X_p  =  \bar{S}_p (\sum_{q \in P \setminus \{p\}} X_q + A) \qquad (p \in P).
\end{equation*}
in $\Inj$. The free $(\mathds{S}_p)_{p \in P}$-multialgebra on $A$ exists iff
$(S^*_{p}A)_{p \in P}$ exists, and is then carried by $\sum_{p \in P}
S^*_{p}A + A$.  All the results of the section then adapt in the
evident way.

\section{Functors and Monads on $\Set$}

In this section we will discuss properties of endofunctors and monads
on $\Set$ needed for the technical development in the next section. 

\begin{theorem}[Trnkov\'a~\cite{Tr}]
  \label{T-Tr}
  For every set functor $H$ there exists a set functor $\trn H$
  preserving finite intersections and agreeing with $H$ on all
  nonempty sets and functions.
\end{theorem}
In fact, Trnkov\'a gave a construction of $\trn H$ as follows: consider
the two subobjects $t, f\colon 1 \to 2$. Their intersection is the
empty function $e: \emptyset \to 1$. Since $\trn H$ must preserve this
intersection it follows that $\trn He$ is injective and forms (not only a
pullback but also) an equalizer of $\trn H t = Ht$ and $\trn H f =
Hf$. Thus $\trn H$ must be defined on $\emptyset$ (and e) as the
equalizer
\[
\xymatrix{
\trn H \emptyset \ar[r]^-{\trn He}
&
\trn H 1 = H1
\ar@<3pt>[r]^-{Ht}
\ar@<-3pt>[r]_-{Hf}
&
H2.
}
\]
Trnkov\'a proved that this defines a set functor preserving
finite intersections. 

\begin{corollary}
  \label{C-Tr}
  The full subcategory of $[\Set, \Set]$ given by all endofunctors
  preserving finite intersections is reflective. 
\end{corollary}
More formally, we have a natural transformation $r\colon H \to \trn H$
such that for any natural transformation $s\colon H  \to K$, where $K$
preservers intersections, there is a unique natural transformation
$\ex s\colon \trn H \to K$ such that $\ex s \o r = s$. 

\begin{IEEEproof}
  From $t \o e = f \o e$ we obtain $Ht \o He = Ht \o He$. Therefore,
  the universal property of the equalizer induces a unique map
  $r_\emptyset\colon H\emptyset \to \trn H\emptyset$ such that 
  $
  He = \trn H e \o r_\emptyset.
  $
  This yields a natural transformation 
  \[
  r\colon H \to \trn H
  \]
  with the component $r_\emptyset$ and with $r_X = \id_{HX}$ for all
  $X \neq \emptyset$. 

  Now let $K$ be an endofunctor preserving finite intersections and
  let $s\colon H \to K$ be any natural transformation. Then $Ke$ is
  the equalizer of $Kt$ and $Kf$, and so we obtain a unique map $\ex
  s_\emptyset$ as displayed below:
  \[
  \xymatrix{
    \trn H \emptyset
    \ar[r]^-{\trn He}
    \ar@{-->}[d]_{\ex s_\emptyset}
    &
    H1
    \ar@<3pt>[r]^-{Ht}
    \ar@<-3pt>[r]_-{Hf}
    \ar[d]^{s_1}
    &
    H2
    \ar[d]^{s_2}
    \\
    K\emptyset
    \ar[r]_-{Ke}
    &
    K1
    \ar@<3pt>[r]^-{Kt}
    \ar@<-3pt>[r]_-{Kf}
    &
    K2
    }
  \]
  Together with $\ex s_X = s_X$ for all $X \neq \emptyset$ this
  defines a natural transformation $\ex s\colon \trn H \to K$ with $\ex s
  \o r = s$. It is now easy to show that $\ex s$ is unique with this
  property. Thus, $r\colon H \to \trn H$ is a reflection as desired.
\end{IEEEproof}

\begin{definition}
\label{D-Tr}
We call the above reflection $\trn H$ of $H$ (which is unique up to unique
natural isomorphism) the \emph{Trnkov\'a closure} of $H$. For a functor
$H$ preserving finite intersections we can always choose $\trn H = H$. 
\end{definition}

\begin{example}
\label{E-Tr}
Let $C_M$ be the constant functor on $M$, and $C^0_M$ its modification
given by $\emp\mapsto \emp$ and $X \mapsto M$ for all $X \neq
\emp$. Then the Trnkov\'a closure of $C^0_M$ is the embedding $r\colon
C^0_M \to C_M$. 
\end{example}

\begin{remark}
  \label{R-monad-hat}
  Trnkov\'a closure extends ``naturally'' to monads: for every monad
  $\boldS = (S,  \eta, \mu)$ there is a unique monad structure on $\trn S$ for
  which $r$ is a monad morphism. We denote this monad by $\trn S$ and
  call it the \emph{Trnkov\'a closure of the monad $\S$}. 
\end{remark}

\begin{notation}
  \label{N-naught}
  For every monad $\S$ on $\Set$ we denote by $\S^0$ its submonad
  agreeing with $\S$ on all nonempty sets (and functions) and with
  $S^0\emp = \emp$.
\end{notation}

\begin{proposition}
  \label{prop:IV.12}
  Every monad $\S$ on $\Set$
fulfils either $\S \cong \trn \S$ or $\S \cong (\trn{\S})^0$.
\end{proposition}
\begin{example}
  \label{E-naught}
  The exception monad 
  \[
  \M_E X = X + E
  \]
  has the submonad $\M_E^0$ (given by $\emp \mapsto \emp$ and
  $X \mapsto X + E$ for all $X \neq \emp$).
\end{example}

\begin{remark}
  \label{rem:IV.14}
  We say that a set functor $H$ \emph{substantially fulfils} some
  property if its Trnkov\'a closure $\trn H$ fulfils it. For example,
  $C_M^0$ is a substantially constant functor. And $\M_E^0$ is a
  substantially exceptional monad. 
\end{remark}

\begin{example}
  \label{E-exc}
  Substantially exceptional monads have a coproduct with every monad
  on $\Set$. This follows for $\M^0_E$ by an argument analogous to
  that of Example~\ref{E-HPP}.
\end{example}

We finish this section by a result of Koubek~\cite{Ko} about behaviours
of set functors on cardinalities. Using similar ideas, we prove an
analogous result for the above endofunctor $\bar S$. 

\begin{proposition}[Koubek \cite{Ko}] 
  \label{P-ess}
  If a set functor $H$ is not substantially constant (see Remark~\ref{rem:IV.14}), then there exists
  a cardinal $\lambda$ with $\card HX \geq \card X$ for all sets $X$
  with cardinality at least $\lambda$.
\end{proposition}

\begin{theorem} \label{T-card} 
  For every consistent monad $\mathds{S}$ on $\Set$
  which is not substantially exceptional there exists an infinite cardinal $\lambda$ with
  \begin{equation*}
    \card \bar{S}X \geq \card X
  \end{equation*}
  for all sets $X$ of cardinality at least $\lambda$.
\end{theorem}
\begin{IEEEproof}[Sketch of proof]
  Since $\S$ is not substantially exceptional, there exists an
  infinite cardinal $\lambda$ such that for every set $X$ of
  cardinality at least $\lambda$ there exists an element $x$
  in $SX$ such that the coproduct embeddings $v_i\colon X \to X \times X$ (a coproduct
  of $X$ copies of $X$) fulfil: $\Sbar v_i(x)$ are pairwise distinct elements.
  Since $X \times X$ is isomorphic to $X$ this proves $\card \Sbar X \geq \card X$. 
\end{IEEEproof}

\section{A Fixpoint Characterization of Coproducts}

In this section we see a remarkable phenomenon, first studied by
Koubek~\cite{Ko}: that many properties of functors and monads on
$\Set$ may be recovered from merely knowing their behaviour on
cardinals.  As we shall see, an instance of this is the existence of
coproducts of monads. Recall that every cardinal $\lambda$ is considered to be
the set of all smaller ordinals. 

\begin{definition}
  By a \emph{fixpoint} of a set functor $H$ is meant a cardinal
  $\lambda$ such that $\card H\lambda = \lambda$. 
\end{definition}

Recall from Remark~\ref{rem:IV.14} that a set functor is \emph{substantially constant} iff its
domain restriction to all nonempty sets is naturally isomorphic to a
constant functor. Analogously for \emph{substantially exceptional monads}.

\begin{proposition}[Trnkov\'a et al.~\cite{TAKR}]
  \label{P-barr}
  A set functor generates a free monad iff it has arbitrarily large
  fixpoints or is substantially constant. 
\end{proposition}

\begin{lemma} \label{L-least} Let $H$ be a set functor with
  arbitrarily large fixpoints. There exists a cardinal $\lambda$
  such that $F_H$ and $H$ have among  larger cardinals the same fixpoints.
\end{lemma}
Next we characterize finitarity of set functors completely via
fixpoints. Recall that a set functor is finitary iff for every set $X$
and every element $x \in HX$ there exists a finite subset $m: Y
\hookrightarrow X$ with $x \in \ran(Hm)$. This is equivalent to $H$
preserving filtered colimits, see~\cite{AP2}.

\begin{lemma} \label{L-disj} 
  Let $n > \alpha$ be infinite cardinals of
  the same cofinality. Then there exists a collection  of more
  than $n$ subsets of $n$ which are almost $\alpha$-disjoint (i.\,e.,
  have cardinality $\alpha$ and the intersection of any distinct pair
  has smaller cardinality).
\end{lemma}
\begin{remark}
  Almost disjoint collections were introduced by Tarski \cite{Ta}. The
  present result can be found in Baumgartner \cite{BM}.
\end{remark}
The proof of the following proposition uses ideas of Koubek in \cite{Ko}.
\begin{theorem} 
  \label{T-finitary}
  Let $H$ be a set functor that is not substantially constant. Then
  $H$ is finitary iff all cardinals from a certain cardinal onwards are
  fixpoints of $H$.
\end{theorem}
\begin{IEEEproof}[Sketch of proof]
  If $H$ is finitary, and $\lambda$ is an upper bound on $\card Hn$,
  $n\in\Nat$, then every cardinal greater or equal to $\lambda$ is a
  fixpoint. Conversely, if $H$ is not finitary, there exists an
  infinite cardinal $\alpha$ and an element $x \in H\alpha$ not
  reachable from smaller cardinals. Then no cardinal $n$ cofinal with
  $\alpha$ is a fixpoint of $H$. To see this, choose an almost
  $\alpha$-disjoint collection as in Lemma~\ref{L-disj} and express it as a
  family of injections $m_i\colon\alpha \to n$.  By using Trnkov\'a
  closure $\trn H$ we see that the elements $Hm_i(x)$ are pairwise distinct. This
  proves $\card Hn > n$.
\end{IEEEproof}
\begin{proposition}
  \label{P-lambda}
  Let $H$ be an accessible set functor that is not substantially
  constant. Then there exists a cardinal $\lambda_0$ such that all
  cardinals $2^\kappa$ with $\kappa \geq \lambda_0$ are fixpoints
  of $H$. 
\end{proposition}
\begin{theorem} 
  \label{T-nec}
  Two consistent monads $\mathds{S}$ and $\mathds{T}$ on $\Set$ have a coproduct
  iff one is substantially exceptional or they have arbitrarily large
  joint fixpoints ($\lambda = \card S\lambda = \card T\lambda$). 
\end{theorem}		
\begin{IEEEproof}
  (1)~Necessity follows from Theorem~\ref{thm:gu}. If both monads
  are not substantially constant, choose a cardinal $\lambda$ that works for
  $\mathds{S}$ as well as $\mathds{T}$ in Theorem \ref{T-card}. For
  every set $A$ of cardinality at least $\lambda$ we choose sets $X
  \cong \bar{S}(Y+A)$ and $Y \cong \bar{T}(X+A)$ and prove that $X$ is
  a joint fixpoint of $\bar{S}$ and $\bar{T}$ of cardinality at
  least $\card A$. The latter is clear from Theorem \ref{T-card}:
  \begin{equation*}
    \card X = \card \bar{S}(Y+A) \geq \card (Y+A) \geq \card A.
  \end{equation*}
  Analogously, $\card Y \geq \card A$. Thus, $X + A \cong
  X$ and $Y + A \cong Y$, from which we conclude
  \begin{equation*}
    X \cong \bar{S}Y \textrm{ and } Y \cong \bar{T}X.
  \end{equation*}
  We have $\card \bar{T} X \geq \card X$ by Theorem \ref{T-card}, and
  another application of Theorem \ref{T-card} yields
  \begin{equation*}
    \card X = \card \bar{S} \bar{T} X \geq \card \bar{T}X,
  \end{equation*}
  thus the cardinal of $X$ is a fixpoint of $\bar{T}$. Then from $Y \cong \bar{T}X$
  we conclude $X \cong Y$ and this yields, by symmetry, a fixpoint of
  $\bar{S}$. Since in $\Set$ we have $SZ = \bar S Z + Z$, it follows
  that also $S$ and $T$ have arbitrarily large joint fixpoints. 

  (2)~Sufficiency. By Example~\ref{E-exc} we need to prove that if $\S$
  and $\T$ are not substantially constant and have arbitrarily large joint fixpoints, then $\S \oplus
  \T$ exists. Due to Theorem~\ref{T-card} $\bar S$ and $\bar T$ have
  arbitrarily large joint fixpoints too. For every set $A$ let $X$
  be an infinite set of cardinality $\card X \geq \card A$ which is a
  fixpoint of $\bar S$ and $\bar T$. Then $X \cong \bar S(X + A)$ and
  $X \cong \bar T(X + A)$ yields a solution of
  Equation~\refeq{eqn:freebitwo}. Consequently, $\S \oplus \T$ exists
  by Proposition~\ref{prop:initchainalg} and Theorem~\ref{thm:gu}.
\end{IEEEproof}
\begin{notation} \label{N-nec}
  $\P$ denotes the power-set monad (i.\,e. the monad of the
  computational effect of non-determinism). And $\P_f$ the
  finite-power-set submonad (of finitely branching non-determinism). 
\end{notation}
\begin{corollary} \label{C-nec2}
  For every consistent monad $\mathds{S}$ on $\Set$ the following conditions are equivalent:
  \begin{enumerate}[(a)]
  \item all coproducts $\mathds{S \oplus T}$ with monads $\mathds{T}$ exist,
  \item $\mathds{S}$ is substantially exceptional,
  \item the coproduct $\mathds{S} \oplus \P$ exists.
  \end{enumerate}
\end{corollary}
Indeed, since $\P$ has no fixpoint, $(c) \rightarrow (a)$ follows
from the above theorem, $(a) \rightarrow (b)$ is Example
\ref{E-exc} and $(b) \rightarrow (c)$ is clear.
\begin{corollary}
  \label{C-nec2f}
  For every monad $\mathds{S}$ on $\Set$ the following conditions are equivalent:
  \begin{enumerate}[(a)]
  \item $\mathds{S}$ has coproducts with all finitary monads,
  \item the functor $S$ generates a free monad,
  \item the coproduct $\mathds{S} \oplus \P_f$ exists.\\
  \end{enumerate}
\end{corollary}		
Indeed $(b) \rightarrow (a)$ follows from Theorems \ref{T-finitary} and
\ref{T-nec} by using Proposition \ref{P-barr}.
		
(a) $\rightarrow$ (c) is obvious, and (c) $\rightarrow$ (b) also
follows from Theorems \ref{T-finitary} and \ref{T-nec}.

\begin{remark}
  In Corollary~\ref{C-nec2} we could use in lieu of $\P$ any monad
  without fixpoints (e.\,g.~the continuation monad). And in
  Corollary~\ref{C-nec2f} in lieu of $\P_f$ we could use any finitary
  monad that is not substantially exceptional (by applying
  Theorem~\ref{T-card}). 
\end{remark}

\begin{corollary}
  \label{C-free}
  Let $\S$ be a consistent monad and $\F_H$ a free monad. Then a
  coproduct $\S \oplus \F_H$ exists iff $S$ and $H$ have arbitrarily
  large joint fixpoints or one of the monads is substantially
  exceptional. 
\end{corollary}
This follows from Theorem~\ref{T-nec} and Lemma~\ref{L-least}.
\begin{corollary} \label{C4} 
  For every finitary monad $\mathds{S}$ on
  $\Set$ all coproducts with free monads exist.
\end{corollary}
\begin{problem}
  Does every accessible monad on $\Set$ have coproducts with all free monads?
\end{problem}

The following result nicely ``complements'' the preceding corollary:

\begin{corollary} 
  \label{C5}
  A monad $\S$ has coproducts with all finitary monads
  iff a free monad on $S$ exists.
\end{corollary}
\begin{example} \label{E2} 
  We present two free monads on $\Set$ whose
  coproduct does not exist. In other words, two set functors $H$ and
  $K$ generating a free monad but such that $H+K$ does not generate
  one. This is a variation on an example, constructed in \cite{KR}
  under the assumption of generalized continuum hypothesis, of a
  non-accessible functor generating a free monad.
		
  Given a class $A$ of cardinal numbers, we can define a functor $\P_A$ on $\Set$ by
  \begin{equation*}
    \P_A X = \{M \subseteq X; \card M \in A \text { or } M = \emptyset\}.
  \end{equation*}
  For every function $f:X \to Y$ put
  \begin{equation*}
    \P_A f(M) =
    \begin{cases}
      f[M] &\text{ if } f \text{ restricted to } M \text{ is injective}\\
      \emptyset &\text{else}\\
    \end{cases}
  \end{equation*}
  Suppose the complement $\bar A = \mathsf{Card} \setminus A$
  contains, for some infinite cardinal $\lambda$, the interval $(
  \lambda, 2^\lambda]$ (of all cardinals $\lambda < \alpha \leq
  2^\lambda$). Then $2^\lambda$ is a fixpoint of $\P_A$:
  \begin{equation*}
    \card \P_A(2^\lambda) \leq \sum_{\alpha \in A, \alpha \leq 2^\lambda}(2^\lambda)^\alpha \leq \sum_{\alpha \leq \lambda} 2^{\alpha \lambda} = 2^\lambda.
  \end{equation*}
  Let $A$ be a class of cardinals such that both $A$ and
  $\bar A$ contain the intervals $(\lambda, 2^\lambda]$ for
  arbitrary large cardinals $\lambda$. Then $\P_A$ and
  $\P_{\overline{A}}$ generate free monads by Theorem
  \ref{C-barr}. However, $\P_A + \P_{\overline{A}}$ has no fixpoints, thus, it does not generate a free monad.
\end{example}
Finally, we can generalize Theorem~\ref{T-nec} to a family of
monads:
\begin{theorem} \label{thm:arbcardcopr}
  A family of consistent monads on $\Set$ has a coproduct iff
  \begin{enumerate}[(1)]
  \item all those monads that are not substantially
    exceptional have arbitrarily large joint fixpoints or
  \item all monads but at most one are substantially exceptional. 
  \end{enumerate}
\end{theorem}


\section{Conclusions}
We have described coproducts of monads on $\Set$. If one of the monads
is inconsistent (i.\,e.~a submonad of the terminal monad), then so is
the coproduct. For consistent monads
we have shown that coproducts of monads on $\Set$ are
well-behaved and can be concretely described:
\begin{enumerate}[(1)]
\item If two consistent monads have a coproduct, then the coproduct
  injections are injective.
\item A consistent monad has coproducts with all monads iff it is
  substantially exceptional (that is, a submonad of an exception
  monad). 
\item Two consistent monads have a coproduct iff they have arbitrarily
  large joint fixpoints or one is substantially exceptional. 
\end{enumerate}
Moreover, for every consistent monad $(S, \eta, \mu)$ we proved that
complements of the unit form an endofunctor $\bar S$ on the category
$\Inj$ of sets and injections.
We used the functor $\bar S$ to present a formula for
coproducts: Consistent monads $\S$ and $\T$ have a coproduct iff for
every set $A$ the recursive equations
\[
X = \bar S(Y + A) \qquad \text{and}\qquad
Y = \bar T(X + A)
\]
have an initial solution $S^* A, T^* A$; the coproduct monad then
sends $A$ to $S^* A + T^* A + A$. This formula was used by Ghani and
Uustalu~\cite{GU} for ideal monads. We also obtain an iterative
construction of the coproduct: $S^*A$ and $T^* A$ are the
colimits of the chains $S^*_i A$ and $T^*_i A$ starting with $\emp$
and given by $S^*_{i+1} A = \bar S(T^*_i A + A)$ and $T^*_{i+1} A = \bar
T(S^*_i A + A)$. This is a substantially easier and clearer construction than
that presented previously by Kelly~\cite{Ke}.

From the above result we derived that the coproduct of finitary monads
is given by the formula $A \mapsto S^*_\omega A + T^*_\omega A + A$,
and that every finitary monad has a coproduct with all free
monads. Coproducts of a monad and a free monad were described by
Hyland, Plotkin and Power~\cite{HPP}, our results imply that a
consistent monad $\S = (S, \eta, \mu)$ has a coproduct with the free
monad on a functor $H$ iff $S$ and $H$ have arbitrarily large joint
fixpoints or $\S$ is substantially exceptional. 

It is an open problem whether every accessible monad has a coproduct
with every free monad.

%
%
\newpage
\bibliographystyle{IEEEtranS}		
\bibliography{own}

\iffull
\clearpage

%
%
\appendix
\comment{ 
\begin{IEEEproof}[Proof for Definition~\ref{D-subf2}(2)]
The cocone property is established inductively, with the induction
step for $i \leqslant j$ given by
\begin{displaymath}
  \xymatrix{
    H^{i+1}0 \ar@{^{(}->}[d] \ar@/_4pc/[dddr]_{\gcone{Y,\phi}{i+1}} \ar@/^1pc/[r]^{h_{i+1,j+1}} \ar[r]_{Hh_{i,j}} & H^{j+1}0 \ar@{_{(}->}[d] \ar@/^2.7pc/[ddd]^{\gcone{Y,\phi}{j+1}} \\
 GH^{i}0 \ar[dr]_{G\gcone{Y,\phi}{i}} \ar[r]_{Gh_{i,j}} & GH^{j}0 \ar[d]^{G\gcone{Y,\phi}{j}} \\
 & GX \ar[d]_{\theta} \\
 & X 
}
\end{displaymath}
\end{IEEEproof}
} 

In order to prove the unicity in Proposition~\ref{prop:recur} we use
the following
\begin{lemma} \label{lemma:galghomchain}
Let $G$ be an endofunctor on $\Set^{I}$, with a subfunctor $H$ on
$\injcat^{I}$. 
\begin{enumerate}[(1)]
\item \label{item:galghomchain} Any $G$-algebra morphism $f\colon A
  \to B$ 
  is a morphism of canonical cocones, i.e.
  \begin{math}
    \xymatrix{
      H^{i}0 \ar[d]_{\gcone{A}{i}} \ar[dr]^{\gcone{B}{i}} & \\
      X \ar[r]_{f} & Y
    }
    \end{math}
  \item Any $H$-$G$-algebra morphism $\morphism{f}{(X,\theta)}{(Y,\phi)}$ is a morphism of canonical cocones, i.e.
    \begin{math}
      \xymatrix{
        H^{i}0 \ar[d]_{\hcone{X,\theta}{i}} \ar[dr]^{\gcone{Y,\phi}{i}} & \\
        X \ar[r]_{f} & Y
      }
    \end{math}
  \end{enumerate}
\end{lemma}
\begin{IEEEproof}
  \begin{enumerate}[(1)]
  \item The inductive step is given by $A = (X, \theta)$ and $B = (Y,
  \phi)$ by
      \begin{displaymath}
        \xymatrix{
          & H^{i+1} 0 \ar@{_{(}->}[r]
          \ar[dl]_*!/^4pt/{\labelstyle\gcone{A}{i+1}}
          \ar@/^{2.4pc}/[rrd]^(0.75){\gcone{B}{i+1}} & GH^{i} 0
          \ar[dl]_*!/^4pt/{\labelstyle G\gcone{A}{i}}
          \ar[d]^{G\gcone{B}{i}} & \\ 
          X \ar@/_1pc/[rrr]_{f} & GX \ar[r]^{Gf} \ar[l]_{\theta} & GY
          \ar[r]^{\phi} & Y 
        }
      \end{displaymath}
  \item The inductive step is given for $A = (X, \theta)$ and $B = (Y,
  \phi)$ by
  \begin{displaymath}
    \xymatrix{
      &  & H^{i+1} 0 \ar@{_{(}->}[r]
      \ar@/_{2pc}/[dll]_(0.75)*!/^2pt/{\labelstyle \hcone{A}{i+1}}
      \ar@/^{2.4pc}/[rrd]^(0.75){\gcone{B}{i+1}}
      \ar[dl]_*!/^4pt/{\labelstyle H\hcone{A}{i}} & GH^{i} 0
      \ar[dl]_*!/^4pt/{\labelstyle G\hcone{A}{i}}
      \ar[d]^{G\gcone{B}{i}} & \\ 
      X \ar@/_1.5pc/[rrrr]_{f} & HX \ar[l]_{\theta} \ar@{^{(}->}[r] & GX \ar[r]^{Gf}  & GY \ar[r]^{\phi} & Y
    }
  \end{displaymath}
\end{enumerate}
\end{IEEEproof}

\begin{IEEEproof}[Proof of Lemma~\ref{lemma:barprops}]
  Let $D : \boldi \longrightarrow \injcat$ be a diagram, where $\boldi$ is a $\lambda$-filtered small category, with colimit $(V,(\mathsf{in}_{i})_{i \in \boldi})$.  Since $S$ is $\lambda$-accessible, $(SV,(S\mathsf{in}_{i})_{i \in \boldi})$ is a colimit of $SD$.  For $x \in \Sbar V$, we have $x \in SV$ so  $x = (S\mathsf{in}_{i})y$ for some $i \in \boldi$ and $y \in SD_i$.  Suppose $y = (\eta^{S}D_{i})z$ for some $z \in D_{i}$.  Then 
  \begin{eqnarray*}
    x & = & (S\mathsf{in}_{i})y \\
    & = &  (S\mathsf{in}_{i})(\eta^{S}D_{i})z \\
    & = & (\eta^{S}V)\mathsf{in}_{i}\,z  
  \end{eqnarray*}
contradicting $x \in \Sbar V$; hence $y \in \Sbar D_{i}$.  We conclude that $(\Sbar V,(\Sbar \mathsf{in}_{i})_{i \in \boldi})$ is a colimit of $\Sbar D$.
\end{IEEEproof}
\begin{IEEEproof}[Proof of Theorem~\ref{thm:initbialg}]
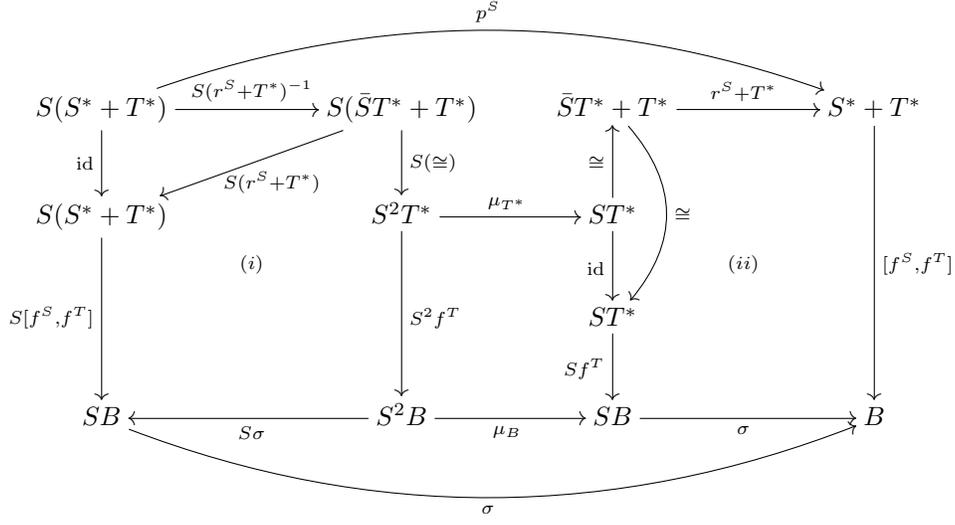
\begin{figure*}[n]
  \begin{displaymath}
    \xymatrix{
      S(S^{*}+T^{*}) \ar@{}[dddrr]|{(i)} \ar@/^{2.5pc}/[rrrrr]^{p^{S}}
      \ar[rr]^{S(r^S + T^{*})^{-1}} \ar[d]_{\id} & & S(\Sbar T^{*} +
      T^{*}) \ar[dll]^{S(r^S + T^{*})} \ar[d]^{S(\cong)} & \Sbar T^{*}
      + T^{*} \ar[rr]^{r^S + T^{*}} \ar@/^{1.7pc}/[dd]^{\cong}
      \ar@{}[rrddd]|{(ii)} & & S^{*} + T^{*} \ar[ddd]^{[f^S,f^T]} \\ 
      S(S^{*} + T^{*}) \ar[dd]_{S[f^S,f^T]}  & & S^{2} T^{*}
      \ar[dd]^{S^{2} f^T} \ar[r]^{\mu_{T^{*}}} & ST^{*} \ar[d]_{\id}
      \ar[u]^{\cong} & & \\ 
      & & & ST^{*} \ar[d]_{Sf^T} & & \\
      SB \ar@/_{2.5pc}/[rrrrr]_{\sigma} & & S^{2} B \ar[ll]^{S\sigma} \ar[r]_{\mu_B} & SB \ar[rr]_{\sigma} & & B 
    }
  \end{displaymath}
  \caption{Showing $[f^S,f^T]$ is an $\boldS$-algebra morphism in the proof of Theorem~\ref{thm:initbialg}(\ref{item:initbialgforw})}
\label{fig:fsftbialghom}
\end{figure*}
For (\ref{item:initbialgforw}) we show that $[f^S,f^T]$ is a an
$\boldS$-algebra homomorphism in Fig.~\ref{fig:fsftbialghom}, and it
is likewise a $\boldT$-algebra homomorphism.  We need only prove part
(ii) of the figure, since (i) is $S$ applied to (ii) and all the other
parts are obvious.  The left-hand component of (ii) is the left-hand
diagram in~\refeq{EQ 3.2} and the right-hand component is given by
\begin{displaymath}
\xymatrix{
 T^* \ar[d]_{\eta^{S}_{T^*}} \ar[dr]^{f^T} & \\
 ST^* \ar[d]_{Sf^T} & B \ar[d]^{\id} \ar[dl]_(0.3){\eta^{S}_B} \\
 SB \ar[r]_{\sigma} & B
}  
\end{displaymath}
For uniqueness, let $g$ be a bialgebra morphism from $(S^* + T^*, p^S,
p^T)$ to $(B,\sigma,\tau)$.  The components of $g$ are $g \o \inl =
f^S$ and $g \o \inr = f^T$. This follows from the commutative diagram below:
\begin{displaymath}
  \xymatrix{
\Sbar T^* \ar[rr]^{r^S} \ar@{^{(}->}[d] \ar[drr]^{\mathsf{inl}} & & S^* \ar@/^2.5pc/[ddd]^(0.2){\mathsf{inl}} \\
ST^* \ar[r]^{\id} \ar[dd]_{S\mathsf{inr}} \ar[dr]^{S\eta_{T^*}} \ar[ddr]_{S\mathsf{inr}} & ST^* \ar[r]_{\cong} &  \Sbar T^* + T^* \ar[dd]_{r^S + T^*}  \\
 & S^2 T^* \ar[u]_{\mu_{T^*}} & \\
S(S^* + T^*) \ar[r]^{S(r^S + T^*)^{-1}} \ar[d]_{Sg} \ar@/_1pc/[rr]_{p^S} & S(\Sbar T^* + T^*) \ar[u]_{S(\cong)} & S^* + T^* \ar[d]^{g} \\
SB \ar[rr]_{\sigma} & & B
}
\end{displaymath}
and the analogous diagram for $g \o \inr$. Indeed, these diagrams
commute since $p^S$ is defined as $(r^S + T^*) \o \mu^S_{T^*} \o S(R^S
+ T^*)^{-1}$, see Remark~\ref{R-trans}, and analogously for $p^T$. 

For (\ref{item:initbialgconv}), assuming an initial bialgebra
$(A,\sigma^{0},\tau^{0})$, we have to show the initial chain
$(S^*_{i}, T^*_{i})$ of (\ref{eqn:initbitwo}) converges. Let
$(\sigma^{0}_{j},\tau^{0}_{j})_{j \in \Ord}$ be the canonical cocone~(Definition~\ref{D-chain})
from the initial chain of (\ref{eqn:initbitwo}) to the
algebra $(A,A,\sigma^0,\tau^0)$ of~(\ref{eqn:initbitwobig}).  If we can
show $\sigma^{0}_{j}$ and $\tau^{0}_{j}$ to be injective for all $j$,
we will be done, as in the proof of
Prop.~\ref{prop:initchainalg}(\ref{item:algchainconv}).



We are going to find, for every
ordinal $i 
$, a bialgebra $(B, \sigma, \tau)$ such that
the canonical cocone $(\sigma_j, \tau_j)$
from the initial chain of (\ref{eqn:initbitwo}) to the algebra
$(B,B,\sigma,\tau)$ of (\ref{eqn:initbitwobig}) fulfils:
\begin{equation*}
  \sigma_i: S^\ast_i \to B \textrm{ and } \tau_i: T^\ast_i \to B \textrm{ are both injective.}
\end{equation*}
This suffices, because the unique bialgebra morphism $h: A\to B$ is
also a morphism of algebras for (\ref{eqn:initbitwobig}), giving by Lemma~\ref{lemma:galghomchain}(\ref{item:galghomchain})
\begin{equation*}
  \sigma_i = h \cdot \sigma^0_i \textrm{ and } \tau_i = h \cdot \tau^0_i
\end{equation*}
which makes $\sigma^0_i$ and $\tau^0_i$ injective.
			\begin{enumerate}[{(b}1)]
				\item We first prove that there exists a $\boldS$-algebra $(B, \sigma)$ of size $\geqslant 2$ and disjoint subobjects
				\begin{equation*}
					s:S^\ast_i \to B \textrm{ and } t:T^\ast_i \to B
				\end{equation*}
				such that the square
				\begin{equation*}
					\xymatrix{
						S^\ast_i \ar[r]^s \ar[d]_{s_{i,i+1}}
						&
						B
						\\
						\bar{S}T^\ast_i \ar@{^{(}->}[d]
						\\
						ST^\ast_i \ar[r]_{St}
						&
						SB \ar[uu]_{\sigma}
					}
				\end{equation*}
				commutes. Here $s_{i,i+1}\colon S^*_i
                                \to S^*_{i+1} = S T_i$ is the
                                connecting morphism of the
                                chain~\refeq{eq:chain1} for $F = \Sbar$ and $G = \Tbar$. Analogously
                                $t_{i,i+1}: T^*_i \to \Tbar S_i^*$.
                                Indeed, let $(B,\sigma)$ the free algebra on
                                $T^\ast_i+2$,  with
				\begin{equation*}
					s:
					\xymatrix@1{
						S^\ast_i \ar[r]^-{s_{i,i+1}}
						&
						\bar{S}T^\ast_i \ar@{^{(}->}[r]
						&
						ST^\ast_i \ar[r]^{S\mathsf{inl}} & S(T^\ast_i + 2)	
					}
				\end{equation*}
				and
				\begin{equation*}
					t:
					\xymatrix@1{
						T^\ast_i \ar[r]^-{\eta^S}
						&
						ST^\ast_i \ar[r]^{S\mathsf{inl}} & S(T^\ast_i + 2)	
					}
				\end{equation*}
				These injections are disjoint by definition of $\Sbar$, and the square commutes due to $\mu^S \cdot S \eta^S = \id$.
				\item For every infinite cardinal
                                  $\kappa \geq \card(ST^\ast_i)$ we can, additionally, require in (b1) that $B$ has cardinality $2^\kappa$. Indeed, starting with an algebra $B_0$ as in (b1), form its power $B = B^\kappa_0$ in $\Set^\mathds{S}$ and take the subobjects $\triangle \cdot s: S^\ast_i \to B$ and $\triangle \cdot t: T^\ast_i \to B$ (for $s$ and $t$ as in (b1)). They are disjoint, and the above square clearly commutes. Since $S^\ast_i$ has at least two elements, so does $B_0 = ST^\ast_i$ due to the injection $s:S^\ast_i \to B_0$. Thus, from $2^\kappa = \kappa^\kappa$ we conclude $\card(B) = \card((ST^\ast_i)^\kappa ) = 2^\kappa$.
				\item By symmetry, given an infinite
                                  cardinal $\kappa$ greater or equal
                                  to the cardinalities of $ST^*_I$ and $TS_i^*$,
				there exists a $\boldT$-algebra $(B', \tau')$ and disjoint subobjects
				\begin{equation*}
					s': S^\ast_i \to B' \textrm{ and } t': T^\ast_i \to B'
				\end{equation*}
				such that the corresponding square commutes and $B'$ has cardinality $2^\kappa$. Since $B \cong B'$ and $s,t$ are also disjoint subobjects, we can find an isomorphism $u:B' \to B$ such that the diagram
				\begin{equation*}
					\xymatrix{
						&
						B' \ar[dd]_u
						\\
						S^\ast_i \ar[ru]^{s'} \ar[rd]_{s}
						&&
						T^\ast_i \ar[lu]_{t'} \ar[ld]^{t}
						\\
						&
						B
					}
				\end{equation*}
				commutes. We let $(B,\tau)$ be the
                                transport of $(B',\tau')$ along $u$
                                (see Remark~\ref{R-trans}). Consequently, the bialgebra $(B,\sigma,\tau)$ has the property that besides the above square also the square
				\begin{equation*}
					\xymatrix{
						T^\ast_i \ar[r]^{t} \ar[d]_{t_{i,i+1}}
						&
						B
						\\
						\bar{T}S^\ast_i \ar@{^{(}->}[d]
						\\
						TS^\ast_i \ar[r]_{Ts}
						&
						TB \ar[uu]_{\tau}
					}
				\end{equation*}
				commutes.
				\item We now prove for all $j \leqslant i$ that
		\begin{equation*}
					\sigma_j = s \cdot s_{j,i} \textrm{ and } \tau_j = t \cdot t_{j,i}
				\end{equation*}
The case $j = i$ implies
				\begin{equation*}
					\sigma_i = s \textrm{ and } \tau_i = t,
				\end{equation*}
				which concludes the proof.
We use induction on $j$, with $j=0$ and the limit case trivial. For the induction step, where $j < i$, we use the following diagram
				\begin{equation*}
					\xymatrix{
						S^\ast_i \ar[rr]^{s} \ar[dd]_{s_{i,i+1}}
						&&
						B
						\\
						&
						S^\ast_{j+1}=\bar{S}T^\ast_{j} \ar[lu]_{s_{j+1,i}} \ar[ru]^{\sigma_{j+1}} \ar[ld]_{\bar{S}t_{j,i}} \ar[dd]^{\phi^S}
						\\
						\bar{S}T^\ast_i \ar@{^{(}->}[dd]
						&
						\\
						&
						ST^\ast_{j} \ar[ld]_{St_{j,i}} \ar[rd]^{S \tau_j}
						\\
						ST^\ast_i \ar[rr]_{S t}
						&&
						SB \ar[uuuu]_{\sigma}						
					}
				\end{equation*}
				(and the corresponding diagram for $t$). It is our task to prove that the upper triangle commutes. Since the outside commutes, see (b1), it is sufficient to observe that all the remaining inner parts commute. For the lower triangle use the induction hypothesis, the right-hand part is the definition of $\sigma_{j+1}$, the left-hand triangle is the definition of $s_{j+1,i+1}$ (as $\bar{S}t_{i,j}$), and the part under it commutes by the naturality of 
                                \begin{math}
\xymatrix{
                                  S' \ar@{^{(}->}[r] & S 
}
                                \end{math}
\end{enumerate}
\end{IEEEproof}

\begin{IEEEproof}[Proof of Lemma~\ref{L-initinj}]
  \begin{figure*}
    \begin{displaymath}
      \xymatrix{
        S({S}^* + {T}^*) \ar[r]^{S(r^{S} + {T}^*)^{-1}}
        \ar@/^3pc/[rrrrr]^{p^{S}} \ar[d]_{\id} & S(\bar{S} {T}^* + {T}^*)
        \ar[r]^-{S(\cong)} \ar[d]^{S(\Sbar g^T + g^T)} \ar[dl]^{S(r^S + T^*)}
        & {S}^2 {T}^* \ar[d]^{S^2 g^T} \ar[r]^{\mu^{S}_{{T}^*}} & S{T}^*
        \ar[r]^-{\cong} \ar[d]^{S g^T} & \bar{S} {T}^* + {T}^* \ar[r]^{r^{S} +
          {T}^* } \ar[d]^{\Sbar g^T + g^T} & {S}^* + {T}^* \ar[dddd]^{g^S +
          g^T} \\ 
        S(S^* + T^*) \ar[dd]_{S(g^S + g^T)} & S(\Sbar {T'}^* + {T'}^*)
        \ar[r]^-{S(\cong)} \ar[d]^{S(\bar{\alpha}_{{T'}^{*}} + {T'}^*)} & S^2
        {T'}^* \ar[r]^{\mu^S_{{T'}^*}} \ar[dd]^{S \alpha_{{T'}^*}} & S{T'}^*
        \ar[r]^-{\cong} \ar[ddd]^{\alpha_{{T'}^*}}  & \Sbar {T'}^* + {T'}^{*}
        \ar[ddd]^{\bar{\alpha}_{{T'}^*} + {T'}^*} & \\ 
        & S(\bar{S'}{T'}^* + {T'}^*) \ar[d]^{\id} \ar[dl]_{S(r^{S'} + {T'}^*)} & & & & \\
        S({S'}^* + {T'}^*) \ar[r]_{S(r^{S'} + {T'}^*)^{-1}} \ar[d]_{\alpha_{({S'}^* + {T'}^*)}} & S(\bar{S'}{T'}^* + {T'}^* ) \ar[r]^-{\cong}
        \ar[d]^{\alpha_{(\bar{S'} {T'}^* + {T'}^*)}} & SS'{T'}^* \ar[d]^{\alpha_{S' {T'}^*}} & & & \\ 
        S'({S'}^* + {T'}^*) \ar[r]^{S'(r^{S'} + {T'}^*)^{-1}}
        \ar@/_3pc/[rrrrr]_{p^{S'}} & S'(\bar{S'} {T'}^* + {T'}^*)
        \ar[r]^-{S'(\cong)} & {S'}^2 {T'}^* \ar[r]^{\mu^{S'}_{{T'}^*}} &
        S'{T'}^* \ar[r]^-{\cong} & \bar{S'} {T'}^* + {T'}^* \ar[r]^-{r^{S'} +
          {T'}^* } & {S'}^* + {T'}^* 
}
\end{displaymath}
                            \caption{Showing $g^{S} + g^{T}$ is an $\boldS$-algebra morphism in the proof of Lemma~\ref{L-initinj}}
                            \label{fig:gsplusgtsalg}
                          \end{figure*}
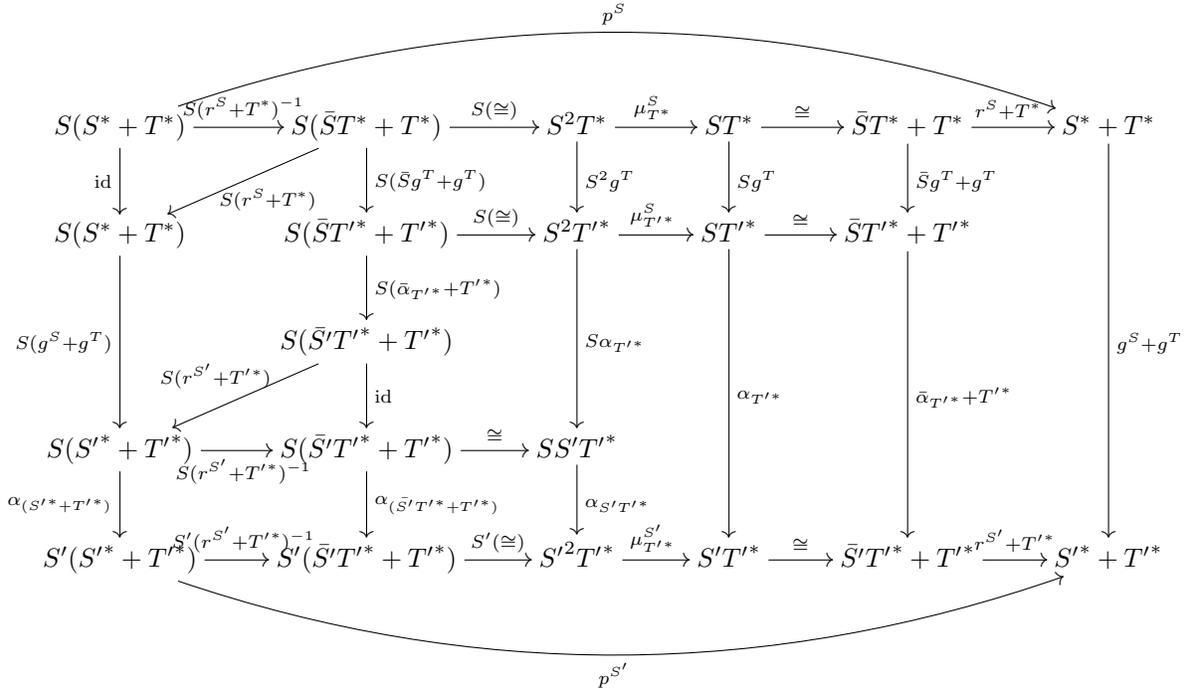
Since $\alpha$ is injective, it restricts to a natural transformation $\bar{\alpha} : \bar{S} \longrightarrow \bar{T}$, and likewise $\beta$ restricts to $\bar{\beta}: \bar{S'} \longrightarrow \bar{T'}$.  By Theorem~\ref{thm:initbialg}(\ref{item:initbialgconv}), the system
\begin{eqnarray*}
    X & = & \bar{S'}Y \\
    Y & = & \bar{T'}X
\end{eqnarray*}
has an initial algebra $((S'^{*},T'^{*}),(r^{S'},r^{T'}))$.  So the system
\begin{eqnarray*}
    X & = & \bar{S}Y \\
    Y & = & \bar{T}X
\end{eqnarray*}
has an algebra
\begin{displaymath}
 P \eqdef (({S'}^{*},{T'}^{*}),(r^{S'}\o\bar{\alpha}_{{T'}^{*}}, r^{T'}\o\bar{\beta}_{{S'}^{*}}))
\end{displaymath}
Therefore, by Prop.~\ref{prop:initchainalg}, it has an initial algebra $((S^{*},T^{*}),r^{S},r^{T})$, and we obtain a unique algebra morphism $(g^{S},g^{T})$ from it to $P$, i.e.
\begin{displaymath}
  \xymatrix{
    \Sbar T^* \ar[rr]^{r^S} \ar[d]^{\Sbar g^S} & & S^* \ar[d]_{g^S} &
    \Tbar S^* \ar[rr]^{r^T} \ar[d]^{\Tbar g^T} & & T^* \ar[d]_{g^T} \\ 
    \Sbar {T'}^* \ar[r]_{\bar{\alpha}_{{T'}^*}} & \bar{S'}T^*
    \ar[r]_{r^{S'}} & {S'}^* &  \Tbar {S'}^* \ar[r]_{\bar{\beta}_{{S'}^*}}
  & \bar{T'}S^* \ar[r]_{r^{T'}} & {T'}^* 
}
\end{displaymath}
Now $S^{*} + T^{*}$ carries an initial $(\boldS,\boldT)$-bialgebra as
described in Theorem~\ref{thm:initbialg}(\ref{item:initbialgforw}).
We show that $g^{S}+g^{T}$ is an $\boldS$-algebra morphism in
Fig.~\ref{fig:gsplusgtsalg} which commutes: recall the definition of
$p^S$ and $p^{S'}$ from Remark~\ref{R-trans} and use the naturality
of $\alpha$ and $\ol \alpha$. Analogously, $g^S + g^T$ for the
exception monad $\M_A$ is likewise a $\boldT$-algebra
morphism.  Therefore it is the desired bialgebra morphism, and it is
injective since $g^S$ and $g^T$ are.
\end{IEEEproof}
\begin{IEEEproof}[Proof of Theorem~\ref{thm:gu}]
The main statement and (\ref{item:guset})--(\ref{item:guunit}) are
immediate from Proposition~\ref{P-freebialg}.  
For the remark: recall that, since $\eta^{S \oplus T} = \inr$, the
coproduct embeddings in Proposition~\ref{prop:algcoprod} are $p^S_A \o
S\inr$ and $p^T_A \o T\inr$, respectively. From the definition of
$p^S_A$ and $p^T_A$, see Remark~\ref{R-trans}, we conclude that the
  diagram in Fig.~\ref{fig:embeddescribe} commutes. And we have an
  analogous diagram from $p^T_A\o T\inr$. This finishes the proof. 
  \comment{ 
$(\boldS \monplus \boldT)m$ is by definition the unique $(\boldS,\boldT)$-algebra morphism
\begin{displaymath}
\morphism{g}{((\boldS \monplus \boldT)A,p^{S}A,p^{T}A)}{((\boldS \monplus \boldT)A,p^{S}A,p^{T}A)}
\end{displaymath}
such that
\begin{displaymath}
  \xymatrix{
  A \ar[r]^{m} \ar[d]_{\eta^{\boldS \monplus \boldT}A} & B \ar[d]_{\eta^{\boldS \monplus \boldT}B} \\
  (\boldS \monplus \boldT) A \ar[r]_{g} &  (\boldS \monplus \boldT) B
}
\end{displaymath}
Clearly $(S^*m + T^*m) + m$ is such a morphism, because everything is
natural. 
 To prove Remark~\ref{???}, we know from Proposition~\ref{prop:algcoprod}(\ref{item:algcoprodforw}) that  the embedding $\morphism{}{\boldS}{\boldS \monplus \boldT}$ is given at $A$ by
\begin{displaymath}
  \xymatrix{
  SA \ar[rr]^-{S \eta^{\boldS \monplus \boldT} A} && S((\boldS \monplus \boldT)A \ar[r]^-{p^{S}A} & (\boldS \monplus \boldT)A
}
\end{displaymath}
and we then apply Fig.~\ref{fig:embeddescribe}.
} 
\begin{figure*}
  \begin{displaymath}
    \xymatrix{
      SA \ar[r]^-{S \inr} \ar[ddd]^(0.3){S \inr}  \ar@/_{3pc}/[ddddd]_{\cong}
      & S((S^*A + T^*A) + A) \ar[d]_{S(\cong)} \ar@/^9.5pc/[ddddddd]^{p^{S}A}  \\
      & S(S^*A + (T^*A + A)) \ar[d]_{S(r^S_A + (T^*A + A))} \\
      & S(\Sbar (T^*A + A) + (T^*A + A)) \ar[d]_{S(\cong)} \\
      S(T^*A + A) \ar@/^3pc/[uur]^-{S\inr} \ar@/^1.5pc/[ur]^-{S\inr} \ar[r]^-{S\eta^{S}_{(T^*A + A)}} \ar@/_1.5pc/[dr]_-{\id} & S^2(T^*A + A) \ar[d]_{\mu^S_{(T^*A + A)}} \\
      & S(T^*A + A) \ar[d]_{\cong} \\
      \Sbar A + A \ar[r]^-{\Sbar \inr  +\inr} \ar[d]_{\Sbar \inr + A} & \Sbar(T^* A + A) + (T^*A + A) \ar[d]_{r^S_A + (T^* A + A)} \\
      \Sbar(T^*A + A) + A \ar[d]_{r^S_A + A} & S^*A + (T^*A + A) \ar[d]_{\cong} \\
      S^*A + A \ar[ur]^-{S^*A + \inr} \ar[r]_-{\inl + A} & (S^*A + T^*A) + A 
    }
  \end{displaymath}
  \caption{Showing embedding description in proof of Theorem~\ref{thm:gu}} 
\label{fig:embeddescribe}
\end{figure*}
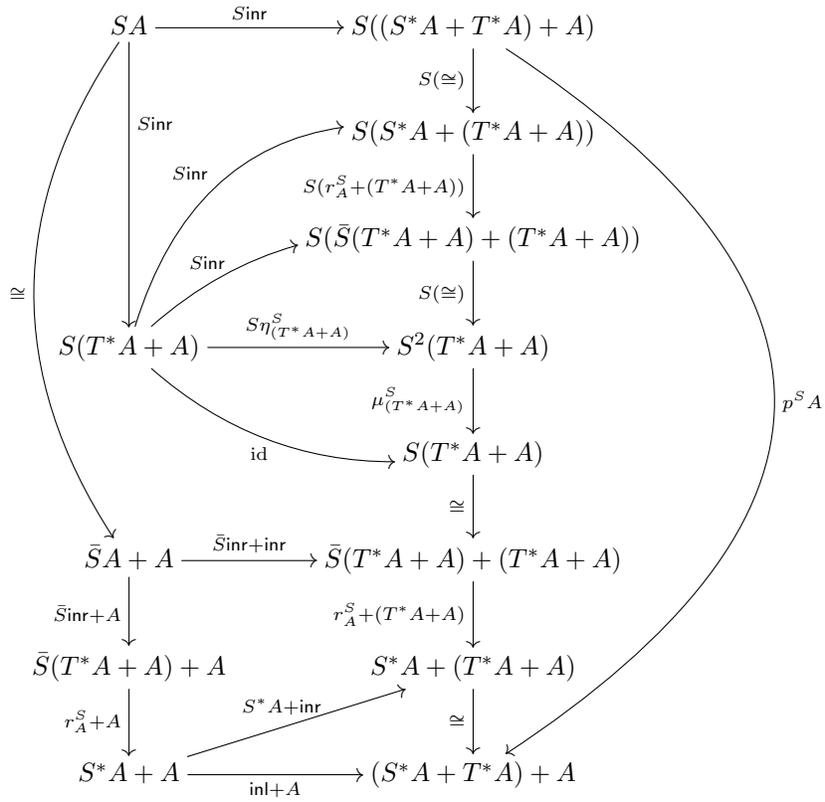
\end{IEEEproof}

\begin{IEEEproof}[Proof of Proposition~\ref{prop:IV.12}]
If $S\emp = \emp$, then $\S \cong \trn\S^0$ follows from the
  fact that $r^S\colon \S \to\trn \S$ has all components on nonempty
  sets invertible. 

  Now suppose that $S\emp \neq \emp$. We want to prove that $r_\emp$ in Corollary~\ref{C-Tr} is
  invertible. Since $e: \emp \to 1$ is injective and $S$ preserves
  injections we conclude from
  \[
  Se = r_1 \o Se = \trn S e \o r_\emp
  \]
  that $r_\emp$ is injective. We will prove that it is a split epic
  by verifying 
  \[
  r_\emp \o \mu_\emp \o \trn S\eta_\emp = \id_{\trn S\emp}.
  \]
  To this end note that $S\emp \neq \emp$ implies $\trn S S\emp = SS\emp$ and $Sr_\emp
  = \trn Sr_\emp$ and consider the diagram below: 
  \[
  \xymatrix{
    SS\emp = \trn S S\emp
    \ar[dd]_{\mu_\emp}
    \ar[r]^-{\trn S r_\emp}
    &
    S\trn S \emp
    \ar@{=}[r]^-{r_{\trn S \emp}}
    &
    \trn S\trn S \emp
    \ar[d]^{\Psi_{S,S}^{-1}}
    \\
    &
    \trn S \emp
    \ar[lu]_-{\trn S\eta_\emp}
    \ar[ru]_{\trn S\trn\eta_\emp}
    \ar@{=}[rd]
    &
    \trn{SS}\emp
    \ar[d]^{\trn\mu_\emp}
    \\
    S\emp
    \ar[rr]_-{r_\emp}
    &&
    \trn S\emp
    }
  \]
  Its outside square commutes since $r$ preserves  multiplication, the
  upper triangle does since $r$ preserves the unit and the 
  right-hand one does by the monad laws of $\trn S$. Thus, the left-hand
  inner part commutes which yields the desired equation.
\end{IEEEproof}

\begin{IEEEproof}[Proof of Theorem~\ref{T-card}]
(a)~We first prove that if a consistent monad $\S$ fulfils $Sf(y) =
  y$ for all endomorphisms $f: Y \to Y$ and all $y \in \Sbar Y$, then
  $\S$ is substantially exceptional. Let $E = S1 \setminus
  \ran(\eta_1)$. We will find a natural isomorphism
  \begin{equation*}
    r_X: X+E \to \mathds{S}X \qquad (\textrm{for all } X \neq \emptyset).
  \end{equation*}
  From that Proposition~\ref{prop:IV.12} implies that $\mathds{S} \cong \mathds{M}_E$
  or $\mathds{M}_E^0$. Given $e \in E$, the element
  \begin{equation*}
    r_X(e) \stackrel{\mathsf{def}}{=} Sg(e) \textrm{ where } g\colon 1 \to X
  \end{equation*}
  is independent of the choice of
  $g$. To see this use the assumption $Sf(y) = y$ for all $y \in \Sbar
  X$ to obtain for every given $g'\colon 1 \to X$ an $f: X \to X$ with
  $g' = f \o g$. This defines the right-hand component of $r_X$, the left-hand
  one is $\eta_X$. Naturality is obvious. The map $r_X$ is injective:
  $\eta_X$ is injective by assumption, $Sg$ is injective because $g$ is a
  split monomorphism, and for every $e \in E$ we have $Sg(e) \notin
  \ran(\eta_X)$ (indeed, $g \cdot h = \id_1$ for $h: X \to 1$, and we
  have $e = Sh(Sg(e)) \notin \ran(\eta_1))$. And $r_X$ is also surjective: for
  every $x \in SX - \eta_X[X]$ apply the above property to the
  endomorphism $f = g \o h$: 
  \begin{equation*}
    x = S \id_X(x) = Sf(Sh(x)) = Sf(e), \textrm{ where } e = Sh(x)
  \end{equation*}

  (b)~To prove the lemma, choose some set $Y$ and an endomorphism $f:Y \to Y$ with
  \begin{equation*}
    Sf(y) \neq y \textrm{ for some } y \in \bar{S}Y.
  \end{equation*}
  Put
  \begin{equation*}
    \lambda = \card Y + \aleph_0.
  \end{equation*}
  
  Given a set $X$ of cardinality at least $\lambda$, there exists $x
  \in \bar{S}X$ such that the coproduct embeddings $v_1, v_2: X \to
  X+X$ fulfil $\bar{S}v_1(x) \neq \bar{S}v_2(x)$; to see this choose
  $m: Y \to X$ and $e: X \to Y$ with $e \cdot m = \id$, and let $x =
  Sm(y)$. We prove the above property by contradiction: Suppose that
  $Sv_1(x) = Sv_2(x)$. Since $g = m \cdot f \cdot e + \id: X + X \to X
  + X$ fulfils $g \cdot v_1 = m \cdot f \cdot e$ and $g \cdot v_2 =
  \id$, thus, $S(mfe)(x) = x$ which, since $x = Sm(y)$, implies that
  \begin{equation*}
    Sm(Sf(y)) = x = Sm(y).
  \end{equation*}
  We know from Lemma~\ref{lemma:monadinj} that $Sm$ is injective, thus, $Sf(y) =
  y$, a contradiction.
		
  We are prepared to prove $\card \bar{S}X \geq \card X$. Since $X$ is
  infinite, we have pairwise disjoint injections $\sigma_i: X \to
  X, i \in I$, where $\card I = \card X$. Arguing as above for $
  \bar{S}v_1(x) \neq \bar{S}v_2(x)$, we see that, for the coproduct
  injections $v_i: X \to \coprod_{i \in I} X$, $\bar{S}v_i(x)$ are
  pairwise distinct for $i \in I$. Since since $\card I = \card X$ we
  have $X \cong \coprod_{i \in I}X$ and therefore $\card \bar{S}X =
  \card \bar{S}(\coprod_{i \in I} X) \geq \card I = \card X$.
\end{IEEEproof}

\begin{IEEEproof}[Proof of Lemma~\ref{L-least}]
  We can assume without loss of generality that $H$ preserves
  injections. (If it does not, use Trnkov\'a closure
  (Definition~\ref{D-Tr}) which has
  essentially the same fixpoints as $H$, and generates a free monad
  iff $H$ does.) Since $H$ is not essentially constant, there exists
  an infinite cardinal $\lambda$ as in Proposition \ref{P-ess}.
			
  We verify that $H$ and $F_H$ have the same fixpoints among sets
  $A$ of at least $\lambda$ elements.
  \begin{enumerate}[(a)]
  \item If $F_H A \cong A$, then $A$ is a fixpoint of $H$ due to $A
    \cong F_HA \cong H(F_HA)+A \cong HA+A$ (see Corollary
    \ref{C-barr}) and $\card HA \geq \card A$ due to the choice of
    $\lambda$.
  \item If $HA \cong A$, then since $A$ is infinite there exists an
    isomorphism
    \begin{equation*}
      a: HA+A \to A
    \end{equation*}
    We define a cocone $f_i:(H+A)^i 0 \to A$ of the initial chain of
    $H+A$, see Definition \ref{D-chain}, by transfinite induction. The
    first step and limit steps are clear. For isolated steps put
    $f_{i+1} = a \cdot (Hf_i + A)$.

    It is easy to see by transfinite induction that all $f_i$'s are
    injective, hence, the free algebra $F_H A$ (which has the form
    $(H+A)^i0$ for some ordinal by Proposition \ref{prop:initchainalg}) has
    cardinality at most $\card (HA+A) = \card A$. Since $F_H A \cong
    HF_H A+A$, we conclude $F_H A \cong A$.
  \end{enumerate}
\end{IEEEproof}

\begin{IEEEproof}[Proof of Lemma~\ref{L-disj}]
  By Zorn's lemma there exists a maximal almost $\alpha$-disjoint
  system $\C$ of subsets of $n$. Assuming $\card \C \leq n$, we derive
  a contradiction. Put $\C = \{X_i; i <n \}$.
			
  Since $n > \alpha$ and $\cof n = \cof \alpha$, there exists a
  strictly increasing sequence of cardinals $n_j$, $j < \alpha$, with
  \begin{equation*}
    \alpha < n_j \textrm{ for all }j \textrm{ and } n = \sup_{j < \alpha} n_j.
  \end{equation*}
  For every $j < \alpha$ we see, since $\card X_i = \alpha < n_j$ that
  $\card\, \bigcup_{i < n_j} X_i \leq n_j$, therefore there exists
  \begin{equation*}
    x_j \in n_{j+1} - \bigcup_{i < n_{j}} X_i.
  \end{equation*}
  The set $A= \{x_j; j < \alpha\}$ meets every member $X_i$ of $\C$ in
  a subset of $\{x_j; j < \alpha$ and $n_j \leq i\}$ which is a set of
  cardinality less than $\alpha$, thus, $\C \cup \{A\}$ is almost
  $\alpha$-disjoint, a contradiction.
\end{IEEEproof}

\begin{IEEEproof}[Proof of Theorem~\ref{T-finitary}]
    Sufficiency. Let $H$ be finitary and $\lambda_0$ be an infinite upper
  bound on the cardinalities of $Hn,\, n<\omega$. For every set $X$ of
  cardinality at least $\lambda_0$ we have, since
  \begin{equation} \label{EQ 2.1}
    HX = \bigcup_{n<\omega} \bigcup_{f:n \to X} \ran(Hf),
  \end{equation}
  that
  \begin{equation*}
    \card HX \leq \sum_{n<\omega} \lambda_0 \cdot \card X^n = \card X.
  \end{equation*}
  Combining this with Proposition~\ref{P-ess} finishes the proof. 

  Necessity. We use the fact that $H$ preserves finite nonempty
  intersections (see Theorem~\ref{T-Tr}.
			
  The equation \refeq{EQ 2.1} characterizes finitary set functors, see
  \cite{AP2}. Suppose that $H$ is non-finitary. Then we can choose
  the smallest cardinal $\alpha$ such that $H \alpha$ is not equal to
  $\bigcup_{n<\omega} \bigcup_{f:n \to \alpha}\ran(Hf)$. It follows
  that
  \begin{equation*}
    H \alpha \neq \bigcup_{\beta < \alpha} \bigcup_{f: \beta \to \alpha} \ran(Hf).
  \end{equation*}
  For otherwise each element of $H \alpha$ is in the range of some
  $Hf$, where $f= \xymatrix@1{n \ar[r]^{f'} & \beta \ar[r]^{f''} &
    \alpha}$, since, by minimality of $\alpha$,
  \begin{equation*}
    H \beta =\bigcup_{n < \omega} \bigcup_{f:n \to \beta} \ran(Hf).
  \end{equation*}			
  We are going to prove that for every set $X$ of cofinality equal to
  that of $\alpha$ we have $\card HX > \card X$. Since there exists
  arbitrarily large such sets $X$, this concludes the proof.
			
  Choose an almost $\alpha$-disjoint family $X_i, i \in I$, as in Lemma
  \ref{L-disj}; thus the index set $I$ fulfils $\card I > \card
  X$. Let $m_i: \alpha \to X$ be the corresponding injections with
  images $X_i$ $(i \in I)$. Without loss of generality $X_i \cap X_j
  \neq \emptyset$ for all $i \neq j$. Choose an element
  \begin{equation} \label{n}
    a \in H \alpha - \bigcup_{\beta < \alpha} \bigcup_{f: \beta \to \alpha} Hf[H \beta].
  \end{equation}
  Then the elements $Hm_i(a)$ are for $i \in I$ pairwise distinct:
  indeed, from $Hm_i(a) = Hm_j(a)$ it follows that $a$ lies in
  $Hf[H(X_i \cap X_j)]$ (where $m_i \cdot f = m_j \cdot g$ is the
  pullback): recall $Hm_i \cap Hm_j = H(m_i \cap m_j)$. This is a
  contradiction because, since $\beta:= \card (X_i \cap X_j) <
  \alpha$, we have that $a$ lies in the right-hand union \refeq{n}
  above. Consequently,
  \begin{equation*}
    \card HX \geq \card I > \card X.
  \end{equation*}
\end{IEEEproof}
\begin{IEEEproof}[Proof of Proposition~\ref{P-lambda}]
  This is analogous to the proof of Theorem~\ref{T-finitary}. Let $H$ be
  $\lambda$-accessible. As proved in~\cite{AP2} this means that the
  formula in~\refeq{EQ 2.1} holds provided that the first union ranges
  over all $n < \lambda$. Let $\lambda_0 \geq \lambda$ be an upper
  bound on cardinalities of $Hn$, $n < \lambda$. For every set $X$ of
  cardinality $2^\kappa$, $\kappa \geq \lambda_0$, we have
  \[
  \begin{array}{rcl}
    \card HX & \leq & \sum\limits_{n < \lambda} \lambda_0 \cdot \card
    X^n \\
    & = & \lambda\cdot\lambda_0 \cdot 2^\kappa \\
    & = & \card X.
  \end{array}
  \]
\end{IEEEproof}
\begin{IEEEproof}[Proof of Corollary~\ref{C4}]
  The statement is trivial in the case where $\S$ is consistent or
  substantially exceptional. So assume that it is not. 
  Let $\lambda$ be a cardinal with $\card SX \geq \card X$ for all
  sets $X$ of cardinality at least $\lambda$, see Theorem
  \ref{T-finitary}. Then either $\mathds{S}$ is not consistent or
  cardinals $\kappa \geq \lambda$ are fixpoints of $S$. For
  every free monad $\mathds{F}_H$ either $H$ has arbitrarily large
  fixpoints, or it is essentially constant, see Proposition
  \ref{P-barr}. In the first case $\mathds{S \oplus F}_H$ exists
  because $HS$ generates a free monad: all fixpoints of $H$ from
  $\lambda$ onwards are fixpoints of $HS$. In the latter case
  $\mathds{F}_H$ is substantially exceptional.
\end{IEEEproof}
\begin{IEEEproof}[Proof of Corollary~\ref{C5}]
  If $\F_S$ exists, then by Lemma~\ref{L-least},
  Proposition~\ref{P-barr} and Theorem~\ref{T-finitary} $S$ has
  arbitrarily large joint fixpoints with every finitary monad. Now
  apply Theorem~\ref{T-nec}. Conversely, if $\S \oplus \T$ exists for
  every finitary monad $\T$, then $S$ has arbitrarily large fixpoints:
  for every cardinal $\lambda$ the monad $\T X = \lambda^* \times X$
  of $\lambda$ unary operations has, by Theorem~\ref{T-nec}, a joint
  fixpoint $\kappa$ with $S$, and clearly $\kappa \geq \lambda$.
\end{IEEEproof}
\fi
\end{document}